# The Evolutionary Dynamics of AI, Politicization, Contestation, and Trust in Science Funding

Animesh Ray

Henry E. Riggs School of Applied Life Sciences

Keck Graduate Institute

535 Watson Drive, Claremont

aray@kgi.edu

&

Division of Biology and Biological Engineering

California Institute of Technology

Pasadena

animesh@caltech.edu

**Significance.** Modern industrial nations rely on public funding of science, certified through expert peer reviews whose authority rest as much on perceived institutional independence as on processing capacity. We model what happens when review or funding decisions are seen as politically directed, and artificial intelligence (AI) substitutes for resistant human reviewers. Treating scientists, the public, and an adaptive funding agency as interacting populations, the model shows that operational capacity and institutional legitimacy obey separate dynamics and can fail independently: automation can sustain a review pipeline while its authority collapses, or legitimacy can recover while backlogs persist. Under finite populations, identical institutions can reach markedly different fates by chance alone, and how often chance favors survival depends on the noise process assumed, not on a fixed institutional probability. This framework identifies which conditions govern whether resistance ignites or whether legitimacy, once threatened, is repaired, as explicit future priorities for empirical calibration and policy design.

**Lay Abstract:** Modern societies depend on public funding for scientific research, a system that only functions if the public believes the process is fair. Using evolutionary game theory, this mathematical work explores what happens when scientific peer review is viewed as politically compromised, and funding agencies turn to artificial intelligence (AI) to replace human experts who protest the system.

The mathematical model reveals several critical warnings for policymakers:

- *Automation Does Not Equal Trust:* Keeping grant money flowing and maintaining public trust are two separate problems. An agency can successfully use AI to process grants while the public completely loses faith in its decisions, creating a "zombie" institution.
- *The Threat of "Proposal Flooding":* When scientists protest, in principle they refuse to undertake peer review, and can overwhelm the funding agency by submitting many proposals. This retaliatory flooding and refusal to review, within a limited budget, can outpace any AI system's ability to keep up, causing the entire pipeline to collapse.
- *Emergency Pay Can Backfire:* Trying to secure scientists' cooperation by raising reviewer pay during a highly politicized crisis can make things worse, as the public may view this emergency pay as a "bribe," which rapidly destroys whatever institutional trust remains.
- *Early Action is Critical:* Human behavior is unpredictable, small early differences in how people react can push identical institutions toward completely different fates. This model tracks individual scientists and members of the public to capture their behavior, which appears to be model-dependent, thus producing sensitivity to early demographic fluctuations and chance factors. Under the standard assumption, every simulated agency ended up with an unmanageable backlog; the only thing chance decided was whether public trust recovered. Under a coarser and more commonly used shortcut, roughly half the agencies appeared to recover fully, and a "zombie" agency that kept processing grants while trust collapsed appeared in about one in six. Policymakers therefore must urgently focus on building trust through transparent, explainable AI rollouts before public rejection becomes permanent.

**Abstract.** Economic stability and progress in modern technological societies are thought to depend on vigorous and independent public funding of science and engineering research. When peer review or funding decisions are perceived as politically directed, scientists, funding agencies, and the public react in coupled and conflicting ways. This manuscript introduces an evolutionary game-theoretic model to analyze how perceived political interference in science funding affects the interrelated behaviors of scientists, funding agencies, and the public. The model simulates scientists choosing to refuse peer reviews or flood agencies with proposals, agencies responding by adopting AI-assisted review and altering reviewer pay, and the public accepting or rejecting these AI systems. Through numerical simulations, five principal findings are identified. (1) Operational capacity and institutional legitimacy are governed by separate conditions and can fail independently. (2) Legitimacy of the process is bistable, meaning final states are determined by the public's acceptance of AI rather than the speed of institutional repair or AI adoption. (3) Since the career cost for researchers refusing to review is generally low relative to politicization, resistance cascades can readily ignite, leading identical institutions to entirely opposite fates based on initial conditions. (4) Increasing reviewer pay only stabilizes participation within a strict budget-solvency frontier; exceeding this boundary destabilizes the system into persistent fluctuations, and emergency pay can paradoxically erode the legitimacy it aims to protect. (5) Finally, finite-population simulations reveal that baseline scenarios partition into either legitimacy recovery without capacity or joint failure, confirming that the fundamental separation of capacity and legitimacy outcomes is a dominant structural feature driven primarily by initial scientific resistance and politicization levels. This theoretical work quantifies issues for future work in science policy.

## Introduction

Public support of scientific research has long been considered a staple of modern democracy, which became a national policy of the US after Vannevar Bush's report to the US President in 1945 (1). In that report it was emphasized that "(the) responsibility for the creation of new scientific knowledge rests on that small body of men and women who understand the fundamental laws of nature and are skilled in the techniques of scientific research…". In the US, the National Science Foundation was established and the research mandates of previously established institutions such as the National Institutes of Health were re-emphasized to catalyze national wellbeing, prosperity and security. Other nations have invested in equivalent programs. One of the important engines of today's over 100-trillion-dollar global economy can ultimately be ascribed to the global government support of research and development, for there are hardly any examples of profound technological progress without crucial pieces of prior art having been discovered by government-funded research. Two critical conditions for the public support of research activity in democracies are (a) the funding of meritorious research proposals, and (b) a sustaining level of public trust on the fairness of government-supported research funding.

Peer review is the mechanism by which public research agencies convert expert judgment by a small body of individuals into authoritative institutional decisions about the scientific merit of proposed research plan. Its output is a ranking as well as a certification of proposals for the soundness and appropriateness of research topics: a signal that scientists, universities, the legal system, and the public are willing to treat as legitimate and as a justification for public investment in scientific and technological research. Historically rooted in the post-WWII social contract of science, when public funding was granted in exchange for rigorous, autonomous self-regulation (1), this mechanism serves as a vital institutional anchor. However, the procedural justice on which this foundation rests is highly sensitive to perceptions of fairness. When peer reviews and/or funding decisions are perceived as politically directed, or when the future introduction of artificial intelligence (AI) agents in peer reviews, currently prohibited by both NIH and NSF peer reviews, threatens to decouple human expertise from perceptions of institutional accountability, the participating actors, namely, the science and engineering community, the public at large, and the funding agencies themselves, react.

Scientists may resist; replacement of human experts by AI-use might increase; the public may distrust the system (2). A significant question in these complex dynamics is whether the resulting decisions made by the funding agencies retain sufficient institutional authority for sustaining public confidence especially when algorithmic efficiency replaces human peer evaluations.

Public trust in governance related to technology adoption has been classically thought to be dependent on two main factors, namely, the perceived ease of use and the perceived usefulness (3). However, there are several fundamental differences between this classical view and the issues in science policy we wish to discuss here. First, the fundamental assumption in the classical view was that public trust in government is generally high in a democracy (3); this is no longer true in a highly politicized and fractured polity (4). Second, classical information technology is transparent in the sense that machine-aided decision-making process can be deconstructed to a set of logical premises and parameters; this is not yet practically feasible with generative AI leveraging deep neural networks (5, 6). Third, and perhaps the most important, the classical model only considered informational machines that were no match for the quality of human cognitive judgment (3); generative AI empowered by large-language models with attention layers is now seen as competitive with human judgement (7–9). These differences force us to consider the potential for further increasing public distrust of government even above the rather high level currently observed (10–12). There is indeed a possibility that above a critical yet currently unknown threshold level of AI adoption by governments, even when such technologies offer both high ease of use and perceived usefulness, trust may instead hinge on a perception of competing interests. Importantly, AI use coupled with the perception of political bias in government funding decisions might lower the threshold of public confidence even further, thus destroying one of the pillars of economic growth of the democratic world.

We model this contest among three actor classes using evolutionary game theory. Scientists decide whether to participate in peer review, a public-goods problem (13), or to flood the system with proposals while boycotting review, shifting the dynamics toward a rent-seeking contest (14). An agency acts as a strategic allocator, adapting its level of AI adoption for reviews and its pay to human reviewers under a finite budget. The public accepts AI review, rejects it, or litigates; AI's effect on trust operates only through AI-review legitimacy, not by assumption. Analyzing this tripartite system through pairwise-comparison imitation dynamics (Fig. 1), we find that at baseline, proposal intensification outruns even maximal AI deployment while trust collapses. Trust recovery depends on the gain in public acceptance, while institutional repair rate sets only how fast a fate is reached, not which one. Resistance by scientists cascades when career cost to experts is low. Reviewer pay can be read as a bribe signal, eroding confidence further. Since these dynamics unfold in finite populations, early demographic chance can carry identical institutions to different fates: how often chance favors recovery is a property of the stochastic process modeled.

## RESULTS

**The Model** To analyze the resilience of the peer-review system under technological disruption, we model the funding ecosystem (see **Appendix I** for a formal description, Supplementary Table S1 for definitions of all state variables, parameters and constants) as an evolutionary game with three interacting entities: an adapting Funding Agency, a population of Scientists, and a population representing the Public actors. This system is driven by co-evolutionary dynamics. The Agency acts as an adaptive controller attempting to clear a backlog of grant proposals while maintaining public trust and managing a finite budget. It does this by continuously adjusting two policy levers: the level of AI adoption for automated peer review (on a continuous normalized scale of 0 to 1), and the financial remuneration offered to human reviewers (likewise normalized from 0 to 1). In response to the Agency's policies, the two strategic populations update their behaviors based on the relative payoffs of their available strategies.

*The Scientist population*: Scientists, who are assumed to be a subpopulation of the general public sphere, dictate the system's human review capacity and proposal load. They choose among four strategies: normal participation, refusing to review (a form of resistance), flooding the system with increased proposal submissions (a form of retaliation, hereby termed ‘proposal intensification’), or both refusing to review and proposal intensification.

*The Public population*: The public dictates the institutional legitimacy of the funding agency (the Agency) and, at large, the prestige of the research enterprise of the nation. Driven by their tolerance for algorithmic decision-making, members of the public choose whether to accept AI-assisted reviews, publicly reject them, or actively litigate to prevent their use. As shown in Fig. 1, we assume that within the non-scientist public sphere there exists an active subset (henceforth termed the Public), which is the effective population considered in the model.

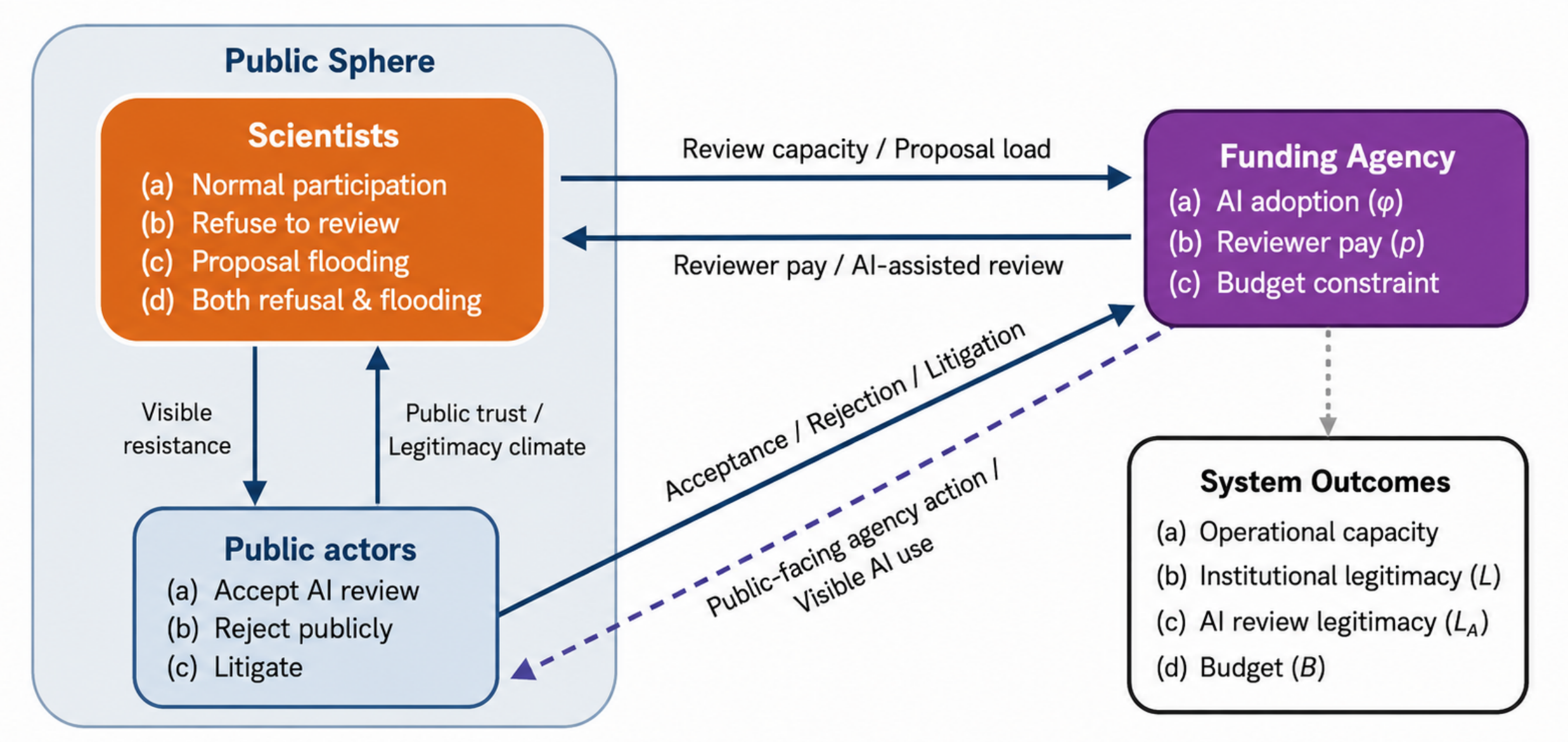

**Figure 1 The peer-review contest: scientists, the public, and an AI-adopting funding agency.** Scientists and Public actors are non-overlapping subpopulations of the general public sphere (Methods): Scientists choose among normal participation, refusal to review, proposal intensification (*i.e.,* 'flooding'), or both; Public actors choose to accept AI-assisted review, reject it publicly, or litigate. The Funding Agency is a capacity-responsive controller rather than an evolving strategic population, setting AI adoption (φ) and reviewer pay (**ρ**) under a finite budget in response to the review capacity and proposal load that Scientist strategies generate. Solid arrows mark flows that enter the payoff or update equations directly; the dashed arrow indicates that the Agency's visible actions (AI deployment, emergency pay) are themselves inputs to Public payoffs, closing the loop between agency behavior and public sentiment. Within the Public Sphere, Scientists' visible resistance and the resulting legitimacy climate are locked in mutual feedback. System Outcomes (bottom right) consist of operational capacity, institutional legitimacy (***L***), AI-review legitimacy (***L****A*), and budget (***B***). All four of the system outcomes are emergent properties of the coupled dynamics.

Thus, the model treats the research-funding system as a co-evolutionary interaction among a population of Scientists, a population of Public actors, and an adaptive Funding Agency. Scientists choose among normal participation ($N$), refusal to review ($R$), proposal intensification ($P$), or the combination of refusal and proposal intensification ($B$). Public actors choose among acceptance of AI-assisted review ($A$), public rejection ($J$), or litigation ($S$). Scientist and Public strategies spread through pairwise-comparison Fermi dynamics (see **Appendix I**) according to their relative payoffs. The Funding Agency is modeled as a capacity-responsive controller rather than as an independently evolving strategic population. This choice reflects, as recent events illustrate, a government funding agency that is practically opaque to the interests of its immediate clients, namely, the scientists. Refusal reduces human review capacity, whereas proposal intensification increases proposal load. From these quantities, the Agency calculates the AI-review deployment required to close the capacity deficit,

$$\phi_t{}^{req} = \text{clip}_{[0,\phi_{max}]}\left[\frac{r_t+\chi p_t}{\beta_{\text{AI}}}\right],$$

where $r_t$ is the fraction of Scientists refusing to review, $p_t$ is the fraction intensifying proposal submission, $\chi$ is the proposal-load multiplier (default: $\chi = 1.5$), and $\beta_{\text{AI}}$ is the review capacity supplied by unit AI deployment. The Agency adjusts actual AI deployment $\phi_t$ toward the required level, subject to fiscal availability and litigation-induced rollback. Reviewer remuneration $\rho_t$ responds separately to scientist resistance.

The interaction of these strategies generates non-linear feedback loops. For instance, if Scientists flood the system with proposals, the resulting congestion forces the Agency to increase AI adoption. This increase in AI reliance may trigger public litigation, which can in reality also drain the Agency's budget (not modeled here) and further damage institutional legitimacy, potentially leading to a collapse of the funding mechanism's authority. Institutional legitimacy $L_t$ and AI-review legitimacy $L_{A,t}$ change through public acceptance, rejection, litigation, expert participation, politicization, and the signaling effects of increases in emergency remuneration as an incentive to the expert-reviewers. The model therefore distinguishes three quantities: review capacity, institutional legitimacy, and AI-review legitimacy.

**Table 1** summarizes the payoff matrices for the Scientists and the Public (derived from equations **1-5**, **Appendix I**). Clearly, $N$ and $P$ retain their career baseline, pay, and legitimacy terms, whereas $R$ and $B$ trade those away to gain the collective-action ($\omega_R$) and the conformity ($\omega_{norm}$) benefits of resisting. The public strategies depend heavily on institutional legitimacy ($L_t$), AI-review legitimacy ($L_{A,t}$), and the exogenous politicization level ($G$).

**In Appendix I**, we derive the replicator dynamics summarized below:

$$\dot{f_i} \approx \left(\frac{\lambda_s}{2}\right) f_i(\pi_i^{(s)} - \bar{\pi}^{(s)})$$

$$\dot{a}_m \approx \left(\frac{\lambda_p}{2}\right) a_m(\pi_m^{(p)} - \bar{\pi}^{(p)})$$

with, $\bar{\pi}^{(s)} = \sum_{i\in \mathcal{S}} f_i\, \pi_i^{(s)}$ and $\bar{\pi}^{(p)} = \sum_{n\in \mathcal{P}} a_m\, \pi_m^{(p)}$, where $\dot{f_i}$ and $\dot{a}_m$ are the rates of change with time of state vectors of strategies employed by the Scientist ($f_i$) and the Public ($a_m$) populations, respectively, $\lambda$ is the corresponding Fermi imitation selection strength (15), and $\pi$ is the corresponding payoff calculated from the payoff matrix (**Table 1**).

**Table 1 Payoff Functions**

| **A** | | *Scientist payoff functions* ($\pi_i(s)$) |
|---|---|---|
| Strategy | Indicator State | Indicator Function ($\pi_i$) |
| $N$ (Normal) | $REV = 1$; $RES = 0$; $PROP = 0$ | $c_0 + \eta\rho_t\omega_{pay} + (-\omega_G G + \omega_L L_t)$ |
| $R$ (Refusal) | $REV = 0$; $RES = 1$; $PROP = 0$ | $c_0 - c_R + \omega_R[r_t - \tau_R]_+ + \omega_{norm} r_t$ |
| $P$ (Flood/Increase) | $REV = 1$; $RES = 0$; $PROP = 1$ | $c_0 + \omega_{prop} + \eta\rho_t\omega_{pay} - \omega_K(K_t - 1)_+ + (-\omega_G G + \omega_L L_t)$ |
| $B$ (Both Refusal + flood) | $REV = 0$; $RES = 1$; $PROP = 1$ | $c_0 + \omega_{prop} - c_R + \omega_R[r_t - \tau_R]_+ + \omega_{norm} r_t - \omega_K(K_t - 1)_+$ |

| **B** | | *Public payoff functions* ($\pi_m(P)$) |
|---|---|---|
| Strategy | Key drivers | Payoff function ($\pi_m$) |
| $A$ (Acceptance) | Favored by high institutional and AI legitimacy | $w_{AL_A} L_{A,t} + w_{A,L} L_t - w_{A,\varphi}(1 - L_{A,t})\varphi_t - w_{A,G} G$ |
| $J$ (Rejection) | Favored by high automation under low legitimacy, politicization, and visible scientist resistance | $w_{J,grow}(1 - L_{A,t})\varphi_t + w_{J,G} G + w_{J,r} r_t - c_J$ |
| $S$ (Litigation) | Favored by public coordination, politicization, and AI adoption, discouraged by individual cost ($c_S$) | $S_t(w_{S,G} G + w_{S,\varphi}\varphi_t) + w_{S,J} y_{J,t} - c_S$ |

These replicator forms are the weak-selection approximations of the full Fermi dynamics (**Appendix I**, **Eqs. 5–6**); all numerical simulations reported below integrate the full Fermi form. Fermi imitation models (15) an actor using strategy $i$ when it compares itself with another actor using strategy $j$, copies $j$ with probability:

$$p(i \leftarrow j) = \sigma\left[\lambda\left(\pi_j - \pi_i\right)\right],$$

where $\sigma(u) = \frac{1}{(1+e^{-u})}$, is the logistic (Fermi) function (16) for $u = \lambda\left(\pi_j - \pi_i\right)$.

**Analytical considerations in Appendix I demonstrate the following important properties of the model:**

(1) The Funding Agency increases AI-assisted review when proposal load $Q_t$ exceeds available human review capacity $H_t$, moving deployment toward the necessary or required level, $\phi_t^{req} = \frac{[Q_t - H_t]_+}{\beta_{\mathrm{AI}}}$, which is equivalent to the expression above because $r_t + \chi p_t = Q_t - H_t$.
Full processing capacity can be restored only when the required deployment does not exceed the technical ceiling $\phi_{\max}$, sufficient fiscal resources are available, and litigation does not force substantial rollback. Because neither institutional legitimacy $L_t$ nor AI-review legitimacy $L_{A,t}$ enters the capacity condition directly, operational capacity and legitimacy remain analytically distinct.

(2) A higher level of politicization of the funding process lowers the critical threshold of resistance by the Scientists. Higher refusal cost, reviewer pay, and institutional legitimacy, all act to raise it. Once resistance by the Scientists exceeds both the threshold of collective action benefits and the critical resistance threshold, refusal to review becomes self-reinforcing under the imitation dynamics.

(3) Public acceptance of the institutional authority is favored under the conditions of high institutional and AI-use legitimacy, but is also favored in regimes of high costs of rejection of institutional authority or litigation by the public. Rejection and litigation are favored by low institutional legitimacy, high automation under low legitimacy, high politicization, visible scientist resistance, and coordinated public rejection.

(4) Institutional legitimacy survives only when repair by acceptance and participation exceeds politicization, rejection, and erosion of legitimacy by the bribe-signal generated by high incentive offered to the Scientists to serve as reviewers.

(5) Importantly, operational capacity and institutional legitimacy can fail independently in either direction. Adequate AI capacity can, in principle, close the review deficit without restoring public acceptance or institutional authority. Conversely, institutional and AI-review legitimacy can recover even when technical, fiscal, or legal constraints prevent the Agency from processing the full proposal load. The model therefore permits capacity-only states, legitimacy-without-capacity states, joint survival, and joint failure.

**Numerical Simulation:** Deterministic trajectories were simulated adaptively, with regime diagnostics applied at 2,000, 5,000, and subsequent 5,000-time-step checkpoints, and a run terminated at the earliest checkpoint at which its regime was classified. The three principal scenarios were followed for a maximum of 50,000 time-steps, whereas parameter-sweep simulations were followed for a maximum of 20,000 time-steps. A trajectory was classified as a fixed point when, over the final 1,000 time-steps, the maximum amplitude across the dynamical state variables was less than $10^{-6}$ and the maximum one-step rate of change was less than $10^{-7}$. Stable cycles were identified only when means and amplitudes remained stable across three consecutive 1,000 time-step windows and a recurrent period was detected; trajectories satisfying neither criterion were classified as unresolved. The baseline and transparent-rollout scenarios satisfied the fixed-point criteria at 20,000 time-steps. Their reported terminal values are therefore fixed-point estimates calculated over the final 1,000 time-steps. Note that the resistance-cascade scenario did not fully satisfy the fixed-point or the stable-cycle criteria after 50,000 time-steps; its reported values are terminal-window means and ranges over the final 1,000 time-steps, and the regime persistently fluctuates within a narrow range. The simulations below locate the attractors of the coupled bounded system, test the threshold relations described above when feed-back interactions operate simultaneously, and show where stochastic finite-population drifts can move trajectories across the basin boundaries. **Supplementary Results Table S2** summarizes the correspondence between analytical threshold conditions (**Appendix I**) and their numerical results.

**Scientists may resist by pressuring the system of peer reviews by intensifying proposal-submission:**

Since the individual benefit of proposal submission, $\omega_{prop}$, is available to Scientists independently of politicization and of refusal to participate in peer reviews, fixation of submission intensification admits more than one behavioral reading: protest against perceived unfairness, ordinary opportunity-seeking, or an AI-enabled fall in the cost of producing a proposal. To test whether the model can distinguish these, we re-specified the submission benefit so that part of it is contingent on politicization and on the prevalence of resistance, giving a genuine collective-action channel parallel to $\omega_R$. The ordinary, explicitly protest-contingent, and mixed specifications all reach the same terminal state ($p^* \rightarrow 1.000$, $K^* = 1.567$, $L^* = L_A{}^*$ = 0). The baseline dynamics therefore do not discriminate among these motivations, and we describe the mechanism in behaviorally neutral terms; the protest (resistance to review and/or proposal intensification) is one interpretation the model permits.

We find that normal participation by scientists falls, and their retaliation against the system is expressed as a rise in proposal submissions rather than as refusal to review (Fig. 2A). Interestingly, the proportion of outright refusers ($R$) declines from the outset, and the fraction pursuing both refusal and intensified submission ($B$) rises only marginally and transiently before it, too, decays; both resistance strategies, together with normal participation, ultimately vanish as proposal intensification ($P$) approaches fixation at the classified fixed point ($p^*$= 1.000). Figure 2A displays the early transients (0-5,000 time-steps), which captures the strategic reversal, alongside the terminal window (19,000–20,000 steps) confirming the fixed point; the terminal values quoted below are fixed-point estimates.

**AI deployment fails to restore review throughput when scientists retaliate by proposal intensification:** Under the baseline parameterization, refusal by scientists to review essentially declined to zero, but proposal intensification became the universal scientist strategy (Fig. 2A). The terminal resistance fraction was only $r^* \approx 0$, whereas the proposal-intensification fraction reached fixation, $p^* = 1.0$. Consequently, human review capacity remained nearly complete, $H^* = 1.0$, but normalized proposal load rose to $Q^* = 2.5$.

The capacity-responsive controller increased AI deployment to its ceiling, $\phi^* = 0.850$, and the budget remained fully available, $B^* = 1.0$. Thus, the failure to clear the proposal load was not caused by fiscal scarcity. Rather, the unconstrained AI requirement rose to $\phi^*_{\mathrm{req,raw}} = 2.143$, well above the allowed ceiling. The resulting effective review capacity was only $C^* = 1.595$, giving terminal congestion $K^* =$ $\frac{Q^*}{C^*} = 1.567$, and a residual capacity deficit of 0.905. Review capacity remained below proposal load throughout the terminal window.

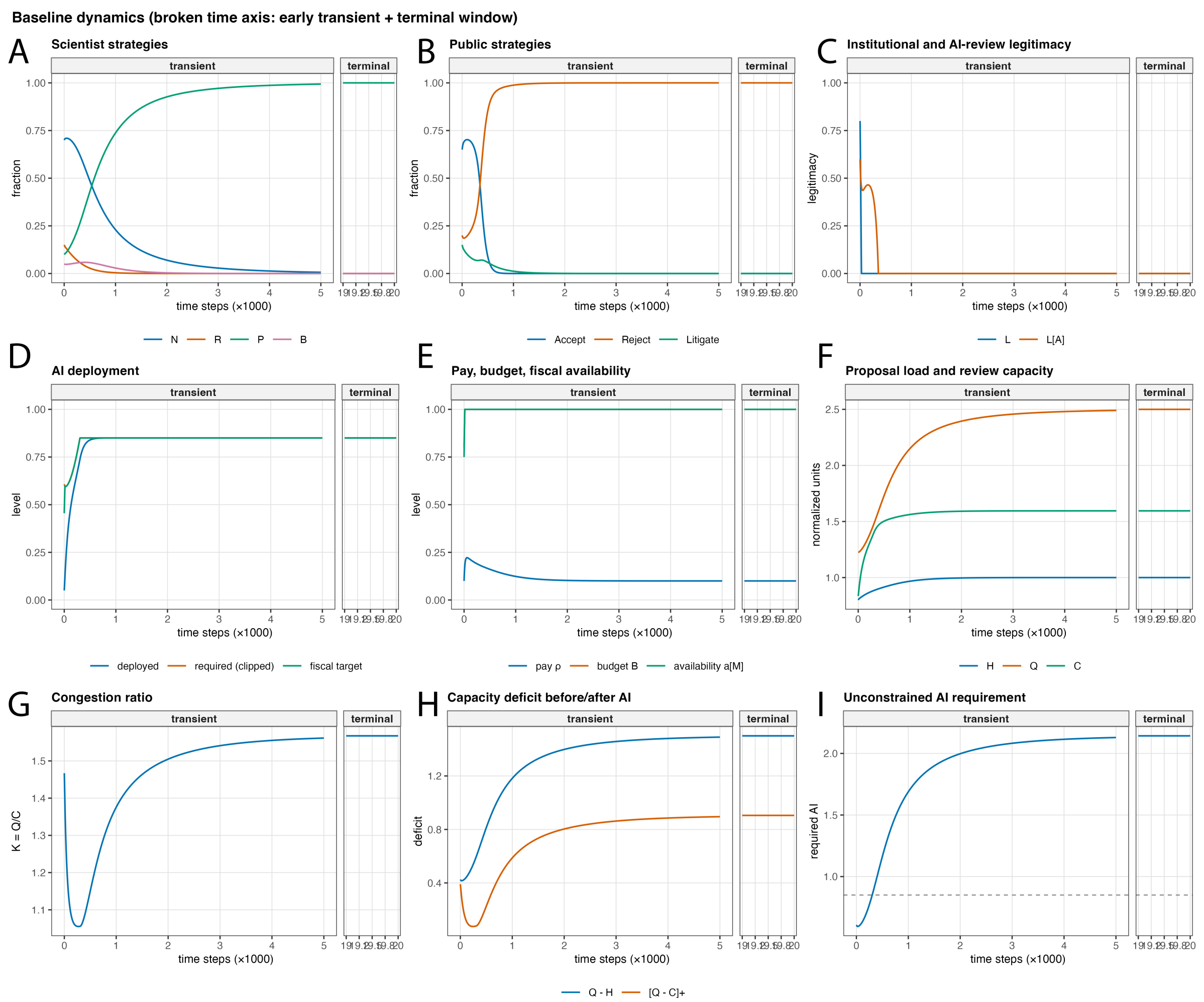

**Figure 2 Baseline evolutionary dynamics: capacity and legitimacy fail together, for different reasons.** Two windows of the same trajectory are shown per panel — an early transient (time steps $\mathbf{0-5,000}$) and the terminal window ($\mathbf{19,000-20,000}$), separated by a break in the time axis. Panels (A, B) Proposal intensification ($\boldsymbol{P}$) reaches fixation ($\boldsymbol{P^*} \approx \mathbf{1.0}$) while normal participation and refusal decay to zero, leaving residual resistance $\boldsymbol{r^*} \approx \mathbf{0.004}$. Public rejection approaches fixation; litigation subsides to near zero, indicating disengagement. Panel (C) Institutional legitimacy ($\boldsymbol{L}$) and AI-review legitimacy ($\boldsymbol{L_A}$) collapse to zero within a few hundred steps. Panels (D, E) AI deployment ($\boldsymbol{\varphi}$) saturates at its ceiling ($\boldsymbol{\varphi_{max}} = \mathbf{0.85}$); budget and fiscal availability remain at maximum throughout, so the ceiling is the binding constraint. Panels (F–I) Proposal load ($Q$) rises to 2.5 against an effective capacity (C) plateau of ≈1.6, producing a residual capacity deficit of ≈0.9 and terminal congestion $\boldsymbol{K^*}$≈1.57. The unconstrained AI requirement (I) rises to ≈2.14, exceeding the deployment ceiling (dashed line): the shortfall is technical.

To confirm that this conclusion is a property of the model itself and not an artifact of parameterization, we constructed a family of counterfactual parameterizations in which $\beta_{AI} \cdot \varphi_{max}$ is set at and above the exact theoretical maximum of the capacity gap $(1 + \chi)$, so that technical infeasibility is structurally impossible. This is verified numerically as the technical shortfall ≡ 0 throughout. Across this entire family, spanning AI productivity from just above the feasibility threshold to five times that margin, the fiscal gap $(\varphi_{req} - \varphi_{target})$ remained substantial (Supplementary Fig. S5). However, deployed AI still fell short of its own fully-feasible requirement, driven entirely by resistance-induced budget stress. The shortfall in AI deployment under resistance is therefore not an artifact of assuming AI is technically underpowered; it is an emergent consequence of the shared-budget feedback between reviewer remuneration and AI operating costs, and it persists regardless of how cheap, capable or efficient AI is assumed to be.

**Public confidence plummets with time:** As the review backlog grows, public acceptance of AI-mediated reviews rises briefly during the initial transient and then collapses, giving way to near-universal public rejection (Fig. 2B). Notably, litigation does not accompany this rejection: the litigation fraction declines monotonically to zero despite the collapse of trust. Rejection without recourse to litigation is consistent with a general malaise—the public disengages rather than contests. The driver of this distrust is a sharp drop in institutional legitimacy and authority (Fig. 2C) as AI is deployed to replace human expert reviewers (Fig. 2D). Human review capacity saturates as refusal vanishes, while proposal load rises toward its asymptote, $Q^* = 2.5$; effective capacity initially outpaces load as AI deployment ramps up, so congestion first falls to a minimum ($K \approx 1.06$) but effective capacity then reaches its ceiling-limited asymptote, $C^* = 1.595$ and congestion climbs to its terminal value, $K^* \approx 1.567$ (Fig. 2E). The system cannot be repaired by increasing remuneration to human peer reviewers (Fig. 2F): pay rises transiently in response to the initial refusal seed and then returns to its baseline value, $\rho^* = 0.10$, because refusal becomes rare, while the budget rises to saturation. The baseline is therefore a joint operational and legitimacy failure: the Agency deploys as much AI review as the model permits, but proposal intensification raises demand faster than rate at which the bounded AI capacity can absorb.

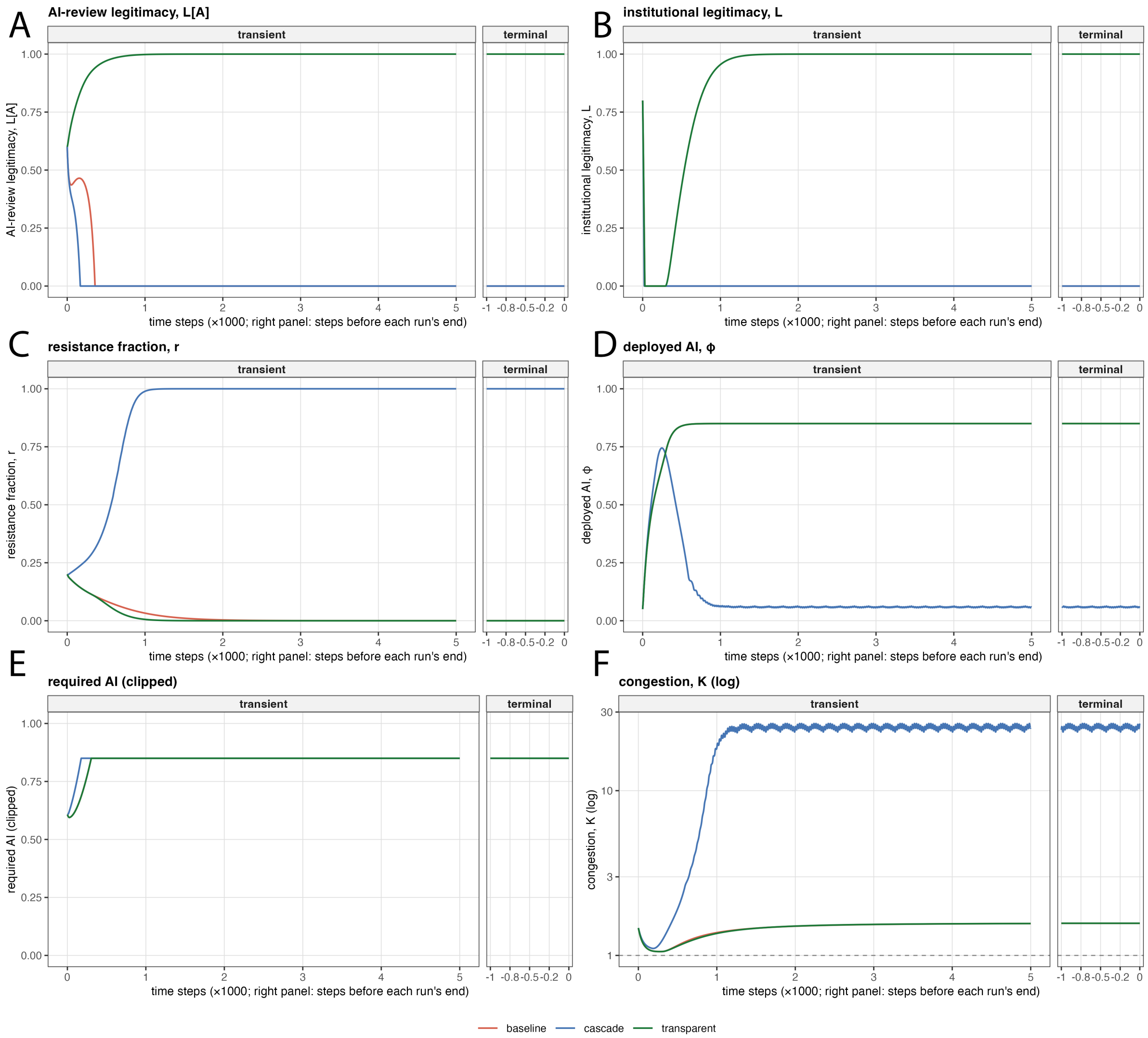

**Figure 3 Legitimacy and capacity fail independently across three scenarios.** Each panel shows two windows of every trajectory: an early transient (0–5,000 steps) and each run's own final 1,000 steps, plotted in relative time since baseline and transparent rollout settle near $t \approx 20{,}000$ while cascade runs the full 50,000 steps without settling. (A, B) Transparent rollout (acceptance gain, $\boldsymbol{\alpha_A = 0.10}$) restores both legitimacy stocks completely ($\boldsymbol{L^* = L_A^* = 1.0}$); baseline and cascade both collapse to zero. (C) Resistance falls to near zero under the baseline and transparent rollout, and reaches fixation ($\boldsymbol{r^* = 1.0}$) under cascade parameters ($\boldsymbol{G = 0.75}$, career cost $\boldsymbol{c_R = 0.22}$). (D, E) Deployed AI reaches the same ceiling ($\boldsymbol{\varphi^* = 0.85}$) under baseline and transparent rollout; under cascade, fiscal scarcity suppresses deployment to $\boldsymbol{\varphi^* \approx 0.059}$. (F) Terminal congestion is identical under baseline and transparent rollout ($\boldsymbol{K^* \approx 1.57}$) and far higher under cascade ($\boldsymbol{K^* \approx 24.2}$, oscillating). Baseline and transparent rollout reach opposite legitimacy outcomes from an identical operational state, showing legitimacy recovery alone does not restore capacity.

**Legitimacy of the review process can recover without operational recovery:** Increasing the rate at which public acceptance repairs AI-review legitimacy produced a qualitatively different legitimacy trajectory without resolving the review-capacity deficit. In this transparent-rollout scenario, identical to the baseline except that each unit of public acceptance rebuilds AI-review legitimacy roughly twice as fast, as might follow from a rollout accompanied by public consultation and explanation. Both legitimacy stocks recovered completely, $L^* = L_A^* = 1.0$, and scientist resistance declined below $10^{-6}$ (Fig. 3A–C). Nevertheless, proposal intensification again reached fixation, $p^* = 1.000$, producing $Q^* = 2.5$. AI deployment again reached its ceiling, $\phi^* = 0.850$, but effective capacity was only $C^* = 1.595$, giving $K^* = 1.567$ and a residual capacity deficit of 0.905 (Fig. 3D–F). Thus, enhanced public acceptance can restore legitimacy fully, but does not move the system to operational survival. The baseline and transparent-rollout scenarios reached identical operational states despite opposite legitimacy outcomes.

A high-politicization, low-career-cost scenario ($G = 0.75, c_R = 0.22$, versus baseline values of 0.55 and 0.30; the resistance-cascade scenario of Fig. 3) produced a third regime. Here resistance by expert scientists approached fixation, $r^* = 1.0$, human review capacity approached zero, and the proposal-intensification fraction also approached zero. Although proposal load therefore returned to its normalized baseline, $Q^* = 1.0$, effective capacity fell to only $C^* = 0.043$, giving a mean terminal congestion of $K^* = 24.2$. The budget declined to $B^* = 0.255$, reducing fiscal availability ($a_M$) to 0.069 and deployed AI review to $\phi^* = 0.059$ even though the clipped requirement sat at the ceiling, $\phi_{req} = 0.850$ (unconstrained requirement 1.43). Here it is fiscal scarcity, not the technical ceiling, that suppresses deployment. Reviewer remuneration remained far above its baseline (terminal-window mean $\rho \approx 0.44$), sustained by universal refusal—the one regime in which the bribe-signal channel of the legitimacy dynamics is active. Congestion, pay, and the budget continued to oscillate within the terminal window; this condition is therefore an unsettled resistance-cascade regime rather than a stationary outcome, and all values quoted here are means over the final 1,000-step window.

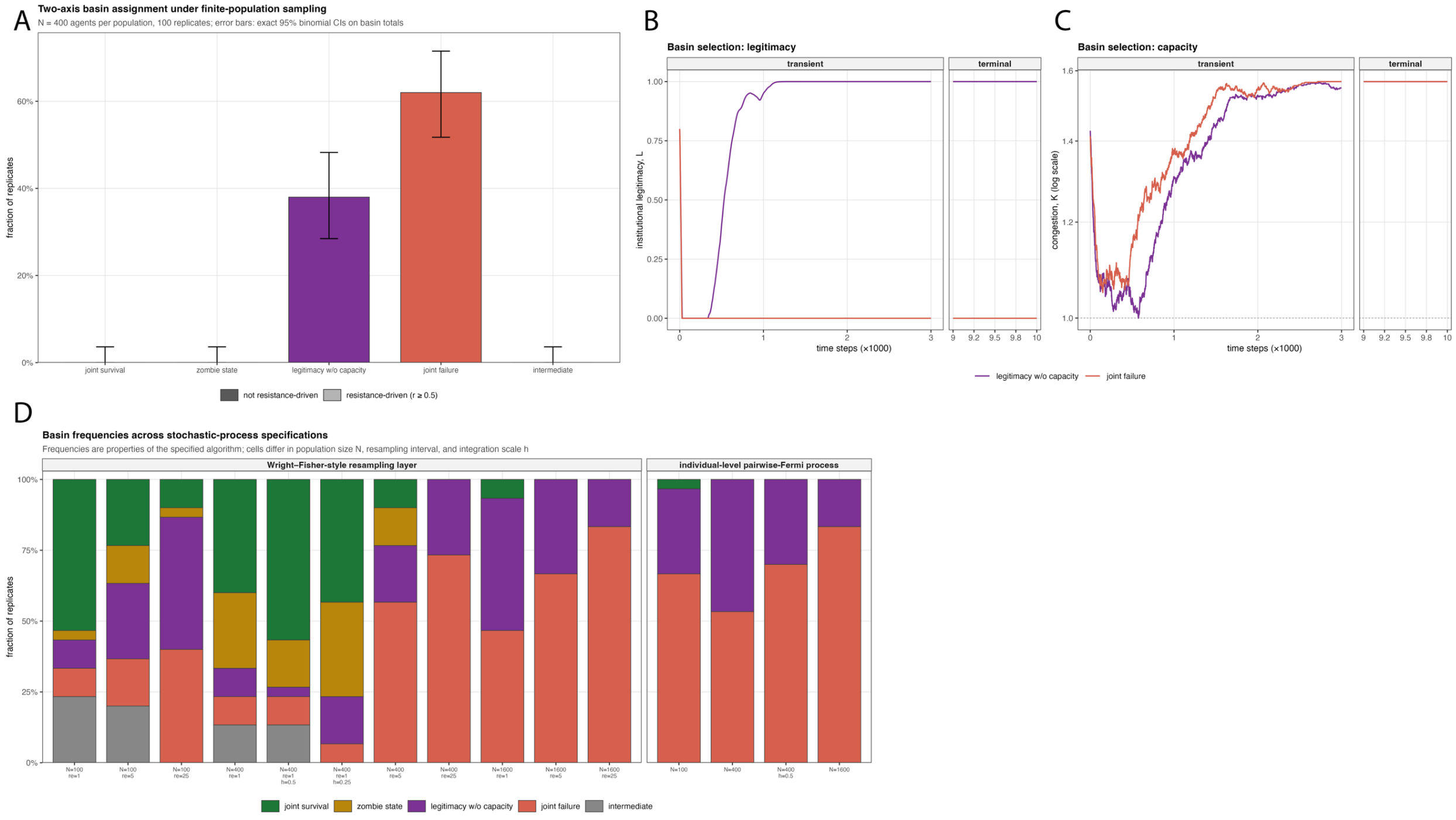

**Figure 4 Basin selection under finite-population sampling is stochastic, mechanistically distinct, and algorithm-dependent.** Panel (A): Terminal basin assignment under a Wright–Fisher-style resampling layer across 100 replicates ($n = 400$ per population, multinomial resampling every step). Shading marks the resistance-driven fraction within each basin ($r \geq 0.5$ in the observation window): joint failure and intermediate are substantially or entirely resistance-driven; the other three basins are not. Panel (B): Terminal basin assignment under the individual-level pairwise-comparison (Fermi) process across 100 replicates launched from the identical baseline initial condition ($n = 400$ per population). Only two of the five outcome classes are realized: legitimacy recovery without capacity (38%, 95% CI 28.5-48.3) and joint failure (62%, 95% CI 51.7-71.5); joint survival, the zombie state, and the intermediate class do not occur (each 0/100, upper 95% bound 3.6%). Panels (C, D): One representative trajectory per basin, launched from the identical baseline initial condition under different noise realizations, on a broken time axis (transient: 0-3,000 steps; terminal: final 1,000 of a 10,000-step run). Congestion (D) separates into three tiers: joint survival and zombie state settle near the operational ceiling ($K \approx 1$); legitimacy without capacity and (non-resistance-driven) joint failure both settle at the deterministic flooding level ($K \approx 1.57$), sharing the baseline's capacity deficit while differing in legitimacy; joint failure's resistance-driven flavor and intermediate settle an order of magnitude higher ($K \approx 20-25$). Legitimacy (Panel C) separates only into recovered (joint survival, legitimacy without capacity) versus collapsed (the remaining three). The two joint-failure modes are indistinguishable in legitimacy and separate in capacity. Panel (E) Basin frequencies under the two stochastic-process algorithms. Under the Wright–Fisher resampling layer, the joint-survival fraction falls as population size $n$ increases; under the individual-level pairwise-comparison (Fermi) process, joint survival is near zero at every tested $n$, consistent with Panel (B). The distribution across joint failure and legitimacy without capacity persists across all specifications tested. However, while the zombie state is a frequent basin in 7 out of 11 populations under the Wright–Fisher sampling, it is never reached under the Fermi process.

**Reachable basins depend on the assumed finite-population process.** Finite-population simulations were performed with 400 Scientists and 400 Public actors under two distinct stochastic-process algorithms. Under a Wright–Fisher-style resampling layer in which the Scientist and Public strategy vectors are resampled as multinomial draws of population size $n$ after each deterministic update, basin assignment across 100 replicates partitions into all five outcome classes (Fig. 4A), namely, joint survival, a capacity-only zombie state, legitimacy recovery without capacity, joint failure, and an intermediate resistance-driven class.

However, under an individual-level pairwise-comparison (Fermi) process, in which each agent independently revises its strategy with a per-step probability proportional to the integration step and copies a randomly drawn co-population member with the standard Fermi probability, only two of these five outcome classes are realized (Fig. 4B): legitimacy recovery without capacity in 38% (95% CI 28.5–48.3) and joint failure in 62% (51.7–71.5). Joint survival, the capacity-only zombie state, and the intermediate class did not occur (each 0/100; upper 95% bound 3.6%), and no replicate was resistance-driven. No replicate switched basins during the 5,000-step observation window.

The two basins realized under the Fermi process are indistinguishable operationally (Fig. 4C, D). In both, resistance vanished ($r^* = 0$), AI deployment saturated at its ceiling ($\varphi^* = 0.850$), congestion settled at $K^* \approx 1.567$ with a residual capacity deficit of $\approx 0.90$, and the budget remained full. They differed only in legitimacy: $L^* = L_A{}^* = 1.0$ in the recovering basin, and $L^* = L_A{}^* = 0$ in the failing one. The capacity-legitimacy separation is therefore realized directly by the stochastic dynamics. Table 2 summarizes the salient results, with 95% confidence intervals determined by nonparametric bootstrap (2,000 resamples) of per-replicate observation-window means (see supplementary files for the remaining results). Values are per-basin means over the 5,000-step observation window across replicates. The replicate distribution is bimodal in legitimacy and unimodal in every operational variable.

**Table 2. Terminal states of basins under the individual-level pairwise-comparison (Fermi) process**

| **Quantity** | **Legitimacy without capacity (mean, 95% cI) ($n$ = 38)** | **Joint failure (mean, 95% cI) ($n$=62)** |
|---|---|---|
| Resistance by Scientists ($r$) | 0 (0) | 0 (0) |
| Institutional legitimacy ($L$) | 1 (0) | 0 (0) |
| Deployed AI ($\varphi$) | 0.85 | 0.85 |
| AI review legitimacy ($L_A$) | 1.00 (0) | 0.00 (0) |
| Budget ($B$) | 1 (0) | 1 (0) |
| Residual capacity deficit | 0.903 (0.901–0.905) | 0.904 (0.903–0.905) |

The two algorithms disagree sharply on which basins are reachable and how their frequencies scale with population size (Fig. 4E). Under the Wright–Fisher resampling layer, the joint-survival fraction falls as population size $n$ increases; under the Fermi process, joint survival is near zero at every tested $n$, consistent with the 100-replicate result in Fig. 4B. The distribution across joint failure and legitimacy without capacity persists across all specifications tested, but the zombie state, a frequent basin under Wright–Fisher sampling (Fig. 4A), is never reached under the Fermi process (Fig. 4B).

**Resistance by Scientists is Guided Primarily by Politicization.** Terminal scientist resistance spanned the entire unit interval across the career-cost-by-politicization plane (Fig. 5A). At low politicization ($G$ = 0.20), resistance vanished at every tested career cost, even when refusal carried no career cost at all, because pay, legitimacy, and career terms jointly outweigh the collective-action and conformity benefits of refusing. By contrast, at high politicization ($G$ = 0.90), resistance remained at fixation ($r^*$ = 1.0) through the cost of resistance $c_R$= 0.33 and then collapsed across a narrow band, falling below $10^{-3}$ by $c_R \approx$ 0.37; on the swept grid the frontier separates full fixation at $c_R$= 0.325 from complete compliance at $c_R$= 0.38. The resulting diagonal frontier agrees with the analytical prediction that politicization lowers the critical resistance threshold whereas the career cost of refusal raises it. The location and sharpness of the frontier were also affected by the Scientist's conviction-driven fraction (Supplementary Fig. S2), collective-action threshold (Supplementary Fig. S2), and selection strength (Supplementary Fig. S2). As expected, strong conviction made remuneration less effective and allowed resistance to persist at larger career costs. Realistically, the career cost to experts who refuse to participate in peer review is minimal in the present system, especially for senior experts. Most of the real-world dynamics should therefore be driven by the extent of politicization.

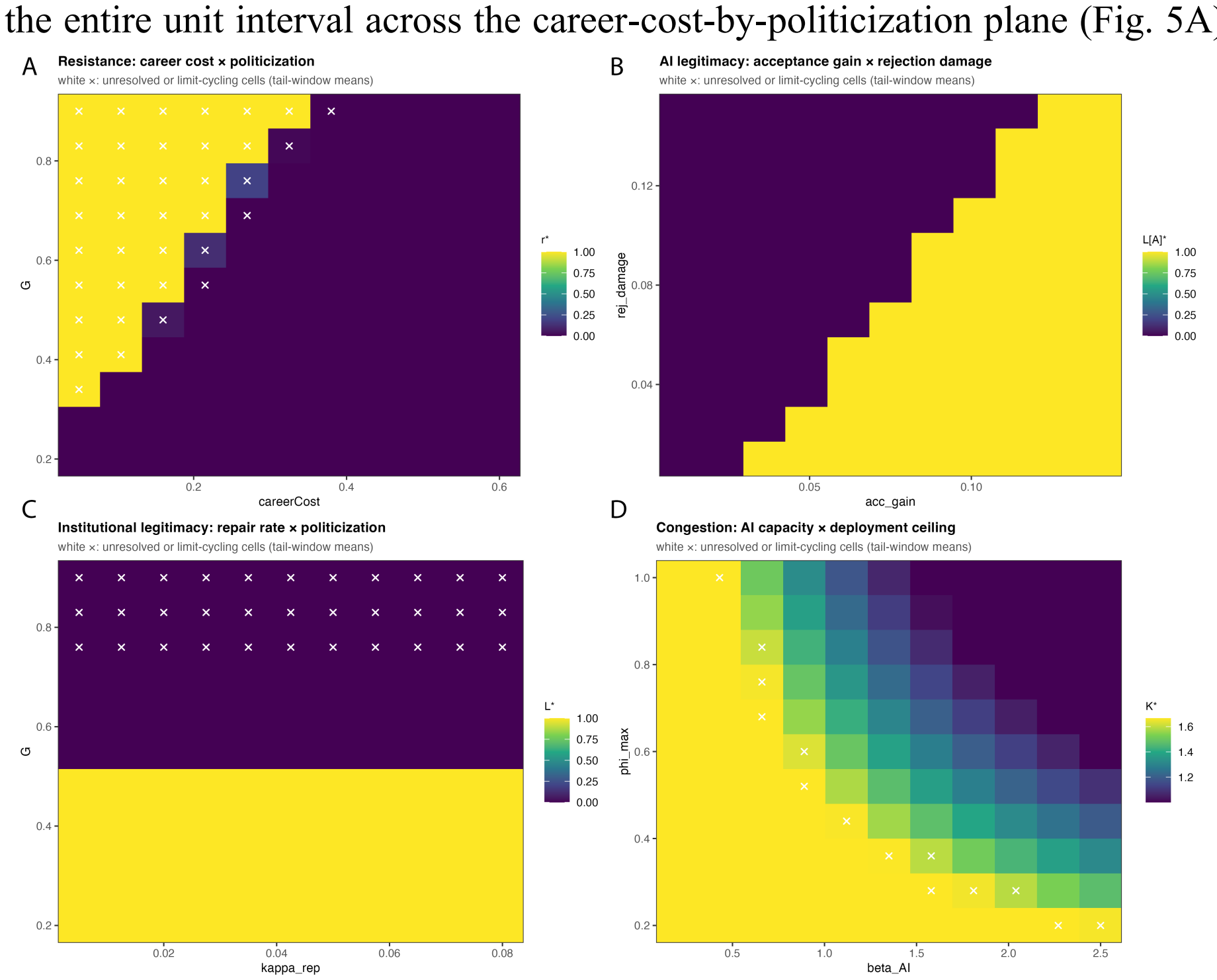

**Figure 5 Threshold structure of resistance, legitimacy, and congestion boundaries.** Each panel is an 11×11 parameter sweep; terminal values are tail-window means, with unresolved or limit-cycling cells marked by white ×. Panel (A) Terminal resistance across career cost ($\boldsymbol{c_R}$) and politicization ($\boldsymbol{G}$) shows a diagonal frontier separating fixation ($\boldsymbol{r^* = 1}$) from compliance ($\boldsymbol{r^* = 0}$); cells near the frontier are frequently unresolved, consistent with the resistance cascade's failure to settle within the sweep horizon. A narrow band of collapse persists within the otherwise-fixed region at high $\boldsymbol{G}$ (≈0.75–0.9): even under heavy politicization, resistance falls back below $10^{-3}$ once career cost exceeds a threshold that itself rises with $\boldsymbol{G}$ (e.g., $\boldsymbol{c_R \approx 0.37}$ at $\boldsymbol{G = 0.90}$), so politicization alone does not guarantee a sustained cascade. Panel (B) Terminal AI-review legitimacy across acceptance gain ($\boldsymbol{\alpha_A}$) and rejection damage ($\boldsymbol{\delta_J}$) shows a sharp, fully-resolved step boundary separating collapse ($\boldsymbol{L_A^* = 0}$) from survival ($\boldsymbol{L_A^* = 1}$); the acceptance gain required for survival rises with rejection damage. Panel (C) Terminal institutional legitimacy across repair rate ($\boldsymbol{\kappa_{rep}}$) and politicization is set by $\boldsymbol{G}$ alone, near $\boldsymbol{G \approx 0.5}$, independent of $\boldsymbol{\kappa_{rep}}$; most cells above the boundary remain unresolved at the sweep horizon. Panel (D) Terminal congestion across the AI capacity multiplier ($\boldsymbol{\beta_{AI}}$) and deployment ceiling ($\boldsymbol{\varphi_{max}}$) declines monotonically with both parameters and remains above the operational threshold $\boldsymbol{K = 1}$ throughout the tested range (in this panel, color scale spans the observed range, ≈1.1–1.7).

Initial conditions further determined which regime was reached. Holding all parameters at baseline, a low initial resistance seed ($r_0$ = 0.07) produced complete legitimacy recovery ($L^*$ = $L_A^*$= 1.0), whereas the intermediate baseline seed ($r_0$ = 0.20) produced legitimacy collapse — despite the two trajectories converging to identical terminal proposal load, AI deployment, and congestion. A high initial resistance seed ($r_0$ = 0.45) instead ignited an unsettled resistance cascade, with complete refusal, deployed AI suppressed by fiscal scarcity, and a depleted budget. The system therefore possesses multiple basins distinguished by both legitimacy and resistance, even though operational capacity is inadequate in more than one of them (Supplementary Fig. S1).

To test whether the capacity–legitimacy separation is an artifact of the model's architecture, we re-ran three scenarios under four alternative specifications that couple the legitimacy and operational subsystems directly: congestion eroding institutional legitimacy, low legitimacy suppressing budget regeneration, low legitimacy constraining AI deployment, and all three simultaneously. The separation persists in every specification. Raising the AI capacity multiplier restores throughput ($K^*$= 1.000) under all five architectures, while raising the acceptance gain restores both legitimacy stocks while leaving the capacity deficit unchanged. This holds even when legitimacy directly gates deployment: under the fully coupled architecture, collapsed legitimacy suppresses deployment to $\varphi^*$ = 0.607 and raises congestion to the flooding-indifference level $K^*$= 1.667, yet the capacity and legitimacy outcomes still vary independently across scenarios.

The baseline scenario is, however, sensitive to architecture in its legitimacy outcome: two couplings move it from joint failure to legitimacy recovery without altering its terminal operational state. Together with the 38/62 replicate split and the wide legitimacy confidence intervals, this indicates that the deterministic baseline lies close to the legitimacy basin boundary — a property we exploit rather than avoid, since it is the regime in which basin selection is informative.

**Fiscal limitation is governed Principally by regeneration and AI operating Cost.** Under the baseline dynamics, refusal to review became rare and reviewer remuneration returned to $\rho_{\text{base}}$. Consequently, changing the coefficient of drain on remuneration over a tenfold range had no discernible effect on the terminal budget at fixed regeneration (Supplementary Fig. S4): Across all tested drain values, the budget returned to $B^* = 1$ for $\delta_{\text{regen}} \geq 0.0144$, and settled at approximately 0.94 at $\delta_{\text{regen}} = 0.0118$, 0.78 at 0.0092, 0.61 at 0.0066, and 0.45 at 0.004, respectively, independently of the drain coefficient. Therefore, it is the budget regeneration, not remuneration drain, that determines the principal fiscal transition. This is consistent with the analytically derived condition for solvency, which at the baseline rest point ($r = 0, \varphi^* = 0.85$) reduces to $\delta_{\text{regen}} \geq q_\phi \varphi^* \approx 0.013$. This lies between the last partially depleted and the first fully solvent grid rows. Cells near the solvency boundary settle slowly or oscillate and are flagged in the figure (indicated by white crosses). A companion sweep of AI operating cost $q_\phi$ against budget regeneration exposed the fiscal tradeoff more directly (Supplementary Fig. S4). Fiscal availability declined continuously as AI operating cost increased and rose with regeneration, ranging from approximately 0.09 to 1 across the tested plane. In the low-resistance baseline regime, AI operating cost is therefore a more consequential fiscal pressure than reviewer-pay drain; remuneration drain becomes important primarily in resistance regimes in which $\rho_t$ remains substantially above its baseline (compare it to the resistance-cascade scenario, where $\rho \approx 0.44$).

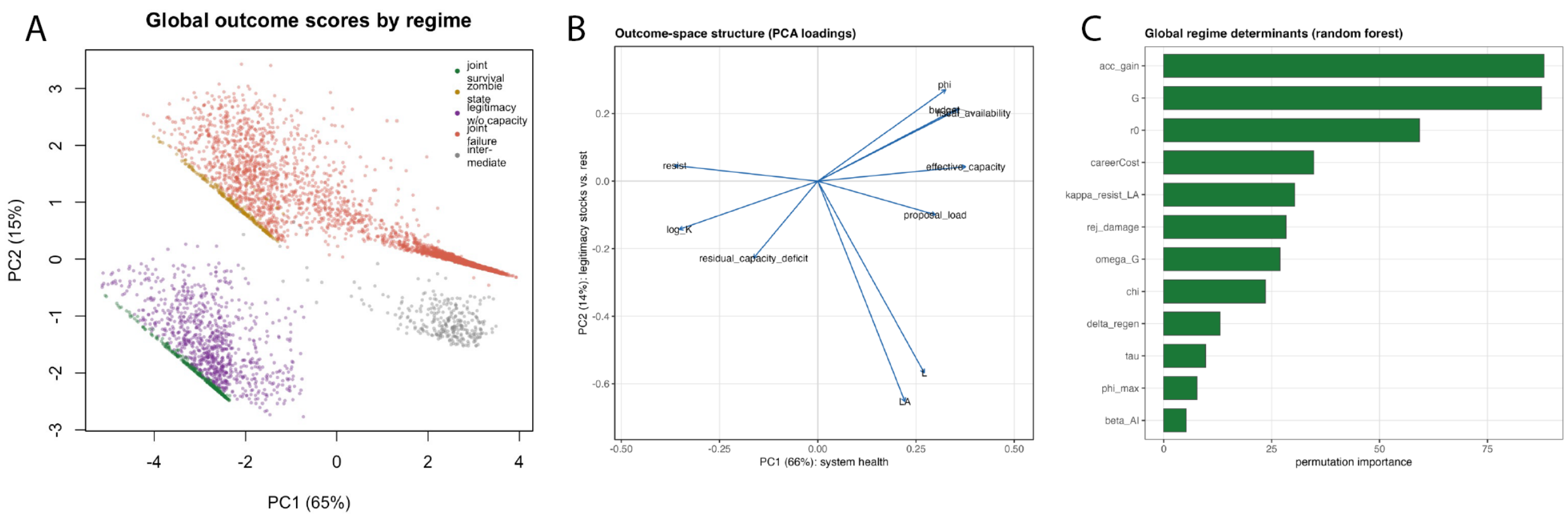

**Figure 6 Global structure of the regime landscape is emergent, not imposed by sampling.** A Latin-hypercube sample of $n$=5,000 draws jointly over all parameters and initial conditions was run to its adaptive terminus and reduced by principal component analysis. Panel (A) Principal Component loadings of the terminal outcome variables. The first component (PC1, 66% of variance) is a system-health axis: budget, effective capacity, fiscal availability, and AI deployment ($\boldsymbol{\varphi}$) load together and oppose resistance and congestion ($\boldsymbol{logK}$). The second component (PC2, 14%) loads almost entirely on the two legitimacy stocks ($\boldsymbol{L}, \boldsymbol{L_A}$), orthogonal to the capacity variables, recovering at the global outcome level the capacity–legitimacy distinction present in the reference architecture. The distinction also persisted under the tested cross-coupled alternatives. Panel (B) Permutation importances from a random-forest classifier predicting the terminal two-axis basin from the input parameters (out-of-bag accuracy 0.742; top 12 of 37 inputs shown). Regime selection is shared between legitimacy-repair parameters. Acceptance gain ($\boldsymbol{\alpha_A}$), rejection damage ($\boldsymbol{\delta_J}$), resistance-driven $\boldsymbol{L_A}$ loss ($\boldsymbol{\kappa_{resist}} \rightarrow \boldsymbol{L_A}$), and the 3 resistance-ignition parameters (politicization ($\boldsymbol{G}$), initial resistance ($\boldsymbol{r_0}$), career cost ($\boldsymbol{c_R}$), consistent with the model's analytic decomposition into two independently governed control problems: which regime is entered, and whether legitimacy survives in that regime.

**Litigation is governed primarily by individual cost and coordination threshold.** Litigation increased as both its individual cost $c_S$ and its coordination threshold $\tau_L$ declined (Supplementary Fig. S4). The response was graded. At the lowest tested litigation cost, $c_S = 0.020$, the terminal litigation fraction declined from 0.967 at $\tau_L = 0.050$ to ~0.28 at the highest thresholds. At $c_S = 0.073$, the corresponding range was approximately 0.74 to 0.14. Litigation was essentially absent when its individual cost exceeded ~0.28, largely independently of the coordination threshold. Litigation affects the coupled system through two distinct channels: it erodes AI-review legitimacy and, when it exceeds the rollback threshold, reduces the deployed AI review. These results describe the strategic conditions under which litigation becomes prevalent. Note, however, that they map the model's assumed incentive structure and are not a prediction of the behavior of the courts.

**Capacity restoration has a sharp phase transition.** Increasing either the AI capacity multiplier $\beta_{\mathrm{AI}}$ or the deployment ceiling $\phi_{\max}$ reduced congestion monotonically (Fig. 5D). Extending the swept domain to $\beta_{\mathrm{AI}} \leq 2.50$ carries the sweep across the analytically derived boundary: terminal congestion spans 1.000 to 1.667, and 20 of 121 cells restore full operational capacity. The transition follows the closed-form condition $\beta_{\mathrm{AI}} \cdot \phi_{\max} \geq \chi$ derived in **Appendix I**, which predicts the simulated capacity classification in 120 of 121 cells; the single exception lies on the boundary contour itself. A companion sweep over $\chi \times \beta_{\mathrm{AI}}$ reproduces the same boundary (119 of 121 cells).

Two features of the transition deserve comment. Beyond the boundary, congestion plateaus at $K = 1$, because the Agency is a sufficiency controller: it deploys only the required level $\varphi_{req}$ and never over-deploys. Below the boundary, congestion is pinned at the flooding-indifference level $K = 1 + \frac{\omega_{prop}}{\omega_K} \approx 1.67$ wherever proposal intensification self-limits at an interior fraction. The location of the boundary is important: at the baseline deployment ceiling $\varphi_{max}$ = 0.85, restoring capacity requires $\beta_{\mathrm{AI}} \geq 1.765$. In other words, AI operating at 85% deployment would have to supply more than one and a half times the review capacity of the entire human reviewer pool. Proposal intensification is the binding constraint throughout the tested baseline-centered domain. The actual limit of AI implementation capacity remains to be evaluated empirically.

**AI-review legitimacy has a sharp acceptance–rejection boundary.** Terminal AI-review legitimacy showed a nearly binary transition across the acceptance-gain by rejection-damage plane (Fig. 5B). Below the transition, $L_A$was driven to zero; above it, $L_A$ approached one. The gain of acceptance by the Public necessary for the survival of institutional legitimacy increased systematically with reputational damage from rejection. For example, when rejection damage was 0.010, survival of legitimacy appeared at a gain of acceptance approximately 0.036, whereas at a rejection damage of 0.150 the transition required an acceptance gain of approximately 0.127. This step-like boundary is consistent with the analytical balance condition requiring legitimacy repair through acceptance to exceed erosion through rejection, litigation, and scientist resistance. A corresponding boundary appeared when acceptance gain was compared with litigation damage (Supplementary Fig. S3). The public cost of rejection also shifted the boundary: high rejection costs reduced the acceptance gain required for AI-review legitimacy to survive, but sufficiently low acceptance gains still produced collapse (Supplementary Fig. S3). Thus, AI adoption rate is not itself the principal driver of legitimacy because AI deployment reached the same ceiling in both the baseline-collapse and transparent-rollout regimes. What mattered was whether the acceptance generated legitimacy rapidly enough to move the coupled public dynamics into the survival basin.

Institutional legitimacy behaved analogously but revealed a stronger form of the same conclusion (Fig. 5C, and Supplementary Fig. S3). Across the repair-rate by politicization plane, terminal $L$ depended only on politicization: below $G \approx 0.5$, $L$ recovered fully at every tested repair rate, and above it, $L$ collapsed at every tested repair rate. In the low-resistance regime, the public converges to a corner of its strategy simplex—full acceptance or full rejection—determined by the politicization-driven public subgame. Once that corner is reached the terminal fate of institutional legitimacy is fixed regardless of how quickly repair operates. The repair rate $\kappa_{rep}$ therefore governs the transient speed of recovery and enters the analytical survival threshold when acceptance is interior (**Appendix I**), but within the tested plane it does not select the terminal basin. The operative lever for both legitimacy stocks is the acceptance dynamics of the public, not the institution's intrinsic repair capacity. Similarly, varying the bribe-signal coefficient did not alter terminal institutional legitimacy in the baseline low-resistance regime, because remuneration returned to its baseline level and the bribe-signal term was therefore inactive (Supplementary Fig. S3).

**Global Structure of the Regime Landscape.** To test whether the regimes identified from targeted scenarios and pairwise sweeps organize the model's behavior globally, we drew a 'Latin-hypercube' sample of $n$ = 5,000 points varying all parameters and initial conditions jointly, ran each to its adaptive terminus, and reduced the terminal-outcome space by principal component analysis (PCA) (Fig. 6A). PC1 (66% of variance) is a system-health axis on which budget, effective capacity, fiscal availability, and AI deployment vary together and oppose resistance and congestion; PC2 (14%) loads almost entirely on the two legitimacy stocks ($L, L_A$), orthogonal to the capacity variables. This separation between capacity and legitimacy was never imposed in the sampling process and is therefore an emergent, model-wide property. A third component (PC3, 12%) loads almost entirely on proposal load and the residual capacity deficit, a raw proposal-pressure axis distinct from both system health and legitimacy. Terminal legitimacy was strongly bimodal across the sample (institutional legitimacy fell below 0.1 in 70% of runs and exceeded 0.9 in 30%), confirming that the basin structure seen in the deterministic scenarios pervades the full parameter space rather than being a feature of the chosen defaults.

A random-forest classifier predicting the terminal two-axis regime from the input parameters achieved 74.2% out-of-bag accuracy, and its permutation importances (Fig. 6B) rank parameters from both control problems comparably: acceptance gain, politicization, initial resistance, and career cost lead, followed by resistance-driven AI-legitimacy loss, rejection damage, and the compliance-politicization weight. These results are consistent with the model's analytic decomposition into two independently governed control problems: career cost, politicization, and initial resistance govern whether a system enters resistance at all, while acceptance gain and rejection damage govern whether legitimacy survives once a regime is entered. Repair rate does not rank among the leading global determinants, consistent with its role as a transient-speed parameter. The two analyses together locate the model's leverage in two distinct places: ignition of Scientist resistance upstream, and legitimacy repair downstream. These results help identify career cost, politicization, initial resistance, acceptance gain, and rejection damage as the priorities for empirical calibration.

## Discussion

Our analytical derivation of the evolutionary game model helps formalize a distinction that is often ignored in policy debates. We demonstrate that substituting automated AI-mediated review for human-expert reviews addresses two classes of inequalities: a capacity inequality and a set of legitimacy inequalities that are mathematically separate. The two can fail independently in either direction. Our simulations realize all four possible combinations. At the deterministic baseline, both fail together. The reason for this failure is instructive: the binding constraint on the throughput of peer-reviews that the funding agency must achieve is not the refusal by human experts to participate in peer reviews in protest, which the imitation dynamics can extinguish. It is in fact the strategy of retaliation by experts who continue to review, thereby avoiding the direct sanction of refusal, while flooding the agency with additional, individually meritorious proposals submitted in protest of perceived unfairness.

We note that the funding agency can in principle limit the number of proposals submitted by an individual Scientist or establish a triage system such as instituting a pre-proposal or a "letter-of-intent" stage to control and/or reduce the load. But that policy is likely to be ineffectual not only because the total number of experts at any one time usually vastly exceeds the number of proposals the agency can handle through human peer reviews but also because the budget that a funding agency has for peer reviews includes both human-expert and machine-aided reviews. This is one counterintuitive feature of the model, that AI deployment never reaches its own technically required level even under sustained resistance and proposal intensification. We emphasize that this is *not* evidence that AI capacity is somehow structurally unable to keep pace with review demand, since it depends entirely on the calibration of the proposal-load elasticity ($\chi$) against maximum AI productivity ($\beta_{AI} \cdot \varphi_{max}$), and would be trivially reversed by assuming cheaper or more capable AI. Instead, we identify and isolate a distinct and more robust mechanism: resistance raises reviewer compensation $\rho^* = \rho_{base} + r_{sci} \cdot (\rho_{max} - \rho_{base})$ as the institution attempts to offset disruption; this elevated pay drains the shared operating budget; and the resulting fall in fiscal availability directly suppresses the controller's own deployment target ($\varphi_{target} = \text{fiscal}_{\text{availability}} \cdot \varphi_{req}$), independent of whether $\varphi_{req}$ is technically achievable at all.

Administrative caps on proposals are already being adopted: NIH has moved to an annual limit of six applications per principal investigator, motivated explicitly by AI-enabled proposal volume rather than by any refusal to review. This is a real-world instance of the submission-intensification mechanism modeled here, and of the policy response the model predicts. It is unclear whether NIH can effectively manage the total number of proposals if every expert submits the maximum allowable amount. While it is true that proposal preparation can take much effort, this can be balanced by engaging AI to construct quality proposals. Thus, concerted action by protesting experts, as modeled here by the Fermi imitation function, could in principle cascade beyond control of the funding agency.

Indeed, in the low-capacity regime, congestion within the funding agencies is pinned at the analytically derived flooding-indifference level $K = 1 + \frac{\omega_{prop}}{\omega_K}$, so that AI substitution cannot restore throughput unless $\beta_{AI}\varphi_{max}$ exceeds the proposal-load multiplier $\chi$. This is a closed-form sufficiency condition for capacity policy (Fig. 5D; Supplementary Fig. S4). Conversely, raising the public acceptance gain restores both legitimacy stocks completely while leaving the identical capacity failure in place. Within finite populations a capacity-only "zombie" state in which its throughput is left intact but its public legitimacy and authority dead, can arise in roughly one replicate in six. Because legitimacy is repaired by acceptance rather than by throughput, restoring review capacity does not restore perceived legitimacy, and vice versa.

Reviewer remuneration can hold participation in human-expert peer reviews steady, but only with sufficient budget, de-politicization, and a small price-insensitive fraction. The budget itself is governed principally by regeneration and AI operating cost, with a solvency frontier that matches the analytical condition $\delta_{regen} \geq \delta_{drain} \cdot r(\rho_{max} - \rho_{base}) + q_{\phi}\varphi^*$. Below this value of $\delta_{regen}$ the fiscal state fluctuates persistently without settling into a demonstrated cycle.

Under high politicization, emergency pay to incentivize peer reviewers can be read as a bribe-signal by public actors, and this perception can damage institutional legitimacy (17, 18), the very property this strategy attempts to preserve. This is an analytical property of the legitimacy dynamics that is inactive in the low-resistance regime where pay never leaves its baseline—a result confirmed by the baseline bribe-signal sweep (Supplementary Fig. S3). It becomes potentially active precisely under the condition where resistance by human experts cascade, where near-universal refusal sustains remuneration far above the baseline. The fiscal solvency frontier and the bribe-signal interaction therefore make incentivizing remuneration a conditional instrument rather than a general remedy. When rejection by the public is high, reactionary response by litigation also collapses to a low level, which can be interpreted as a state of general malaise or public apathy.

Since the legitimacy of AI-review turns on the balance of acceptance against rejection and litigation (Fig. 5B-C), the practical leverage in the model lies in raising public acceptance gains rather than in the speed or extent of AI adoption, which reached the same ceiling in the collapse and recovery regimes alike. Notably, within the tested low-resistance plane, even the institution's intrinsic repair rate did not select the terminal outcome. Terminal legitimacy was determined entirely by the politicization-driven public-acceptance basin. Thus, repair capacity operates as a transient accelerator and an analytic threshold rather than a leverage for basin-selection. These mechanisms are of the kind implicated in recent episodes of rapid automation in federal functions (19) accompanied by significant erosion of trust in government among sections of the public (20), though the model is not calibrated to them.

Two further features of the model bear on such episodes. First, outcomes are strongly initial-condition-, noise- and model-dependent: identical institutions at identical parameters were carried by sampling drift alone into one of several outcomes: joint survival, a capacity-only zombie state, legitimacy recovery without capacity, joint failure, or an intermediate class entirely resistance-driven, with basin assignment completed within the first few hundred steps and no switching thereafter during the 5,000-step observation window. This split reflects one specific finite-population algorithm and population size. However, joint survival and the zombie state both diminish toward negligible levels as population size grows. Under an individual-level pairwise-comparison process, the standard micro-foundation for these dynamics, joint survival and the zombie state (Fig. 4A) are absent, while joint failure and legitimacy-without-capacity persist across every specification tested (Fig. 4B, E). The early phase of a legitimacy dispute is therefore decisive, and interventions that shift even a modest fraction of actors toward acceptance during that window can tip the system into the survival basin. After basins consolidate, the same interventions merely accelerate a fate that is already chosen. Second, this unpredictability is quantifiable. Near the boundary, the distribution of institutional fates is the object any rational policy should target.

Three future directions would be especially valuable: (a) survey-based estimates of price-sensitivity of experts under politicized versus neutral conditions; (b) longitudinal estimates of trust recovery after automated review is introduced; and (c) historical data on participation thresholds in boycotts, open letters, resignations, and litigation coalitions. These map directly onto the parameters this work flags as highest-leverage and least-identified. The model abstracts from field and agency heterogeneity, legal doctrine, appropriations politics, journal-level review, professional societies, and the network structure of scientific communities. The agency is a single adaptive controller, and AI quality is a scalar standing in for accuracy, bias, explainability, confidentiality, and appealability. While the qualitative trajectories identified by our theoretical results are robust, the exact phase-space locations of the cascades, solvency thresholds, and legitimacy boundaries should be treated as calibration targets. Furthermore, the mean-field, well-mixed imitation assumption — already partially relaxed here via an explicit individual-level pairwise-comparison process (Fig. 4C) — must eventually be tested against structured network dynamics, potentially utilizing a Moran process (21–23), within empirically observed social networks of scientists and/or collaborators (24–27).

The work presented here stands on the fundamental assumption that the public is generally aware of the potential impact of a foundational research enterprise free of political interference as a national priority. Estimates of impact of fundamental research vary widely, from an elegant estimate of the return of investment in fundamental research all the way from the time of Galileo Galilei to 1972 as less than equivalent to a mere 12-days-worth of industrial production in the United States in that year (28), to estimates of economic multipliers that vary over a wide range (29–33). Whatever be the actual return, there is little doubt that scientific research, especially fundamental research, is worthy of public support. In this age of generative AI and increasing political polarization, conversations on how, not whether science is funded should take place under a rigorously quantitative framework. This work is an attempt to contribute to that goal.

In summary, our theoretical model suggests that when peer review is contested, an institution's survival depends jointly on capacity and on legitimacy, and the two can fail independently. AI substitution and reviewer pay can together address institutional capacity for moving proposals to funding decisions, but institutional legitimacy depends on public acceptance of AI review despite its potentially compromised quality, on the coordination of resistance and litigation among the public and scientists, and, for how quickly, if not whether trust returns on repair of the reputation of the review process. Under plausible conditions the review pipeline continues while trust in its decisions collapses, under other conditions trust recovers. This model provides a structured framework that maps these mechanisms and the regimes they generate, with explicit parameters that future empirical work can measure. The AI revolution presents a critical stress test for the research-funding establishment. As our model demonstrates, substituting human expertise with automated systems risks unraveling the institutional legitimacy of science funding, requiring an active renegotiation of the social contract between funding agencies, scientists, and the public.

# Methods

### Numerical integration and population dynamics

The numerical implementation used a bounded Euler step followed by renormalization (See Table S1 for definition of terms and **Appendix 1** for analytical derivation of dynamical equations and payoff matrices):

$$x_{t+1} = \mathrm{renorm}(x_t + h_s \dot{x}_t),$$

$$y_{t+1} = renorm\big(y_t + h_p \dot{y}_t\big),$$

where $h_s$ and $h_p$ are the scientist and public step sizes, respectively.

Thus, the simulation rule involves bounded discretization of a pairwise-comparison dynamic. A boundary between the regimes occurs where the relevant payoff difference $\Delta\pi$ changes sign.

### Agency state update rules

*Agency Adaptation:*

At each step the model advances in two stages. First, the Scientist and Public strategy frequencies update simultaneously from the shared time-*t* state (each stratum's Fermi flow reads the time-*t* agency variables $L_t$, $L_{A,t}$, $\varphi_t$, $\rho_t$, and $B_t$ and its own current frequencies). Second, the Agency updates its five scalars from the just-revised frequencies $x_{t+1}, y_{t+1}$ together with its own time-*t* values. Thus, the agency responds to the current step's population movement within the same step, while the populations respond to the agency with a one-step lag (a Gauss–Seidel ordering).

Institutional legitimacy $L$ at a time $(t+1)$ is repaired by public acceptance combined with nonresistant Scientist participation, and is eroded by politicized disruption, rejection of legitimacy by the Public, and the bribe-signal effect of an emergency pay increase above baseline offered as an incentive to review:

$$L_{t+1} = \mathrm{clip}_{[0,1]}[L_t + \kappa_{rep}(1 - L_t)a_{A,t}(1 - r_t) - \kappa_{pol}D_t G - \kappa_{rej}a_{J,t} - \beta_{bribe}[\rho_t - \rho_{base}]_+ G],$$

where $D_t = \mathrm{clip}_{[0,1]}(r_t + a_{J,t} + a_{S,t})$ aggregates the three sources of disruption. AI-review legitimacy $L_A$ is repaired by acceptance (with a saturating $(1 - L_A)$ gain) and eroded by rejection, litigation, and scientist resistance:

$$L_{A,t+1} = \mathrm{clip}_{[0,1]}[L_{A,t} + \alpha_A a_{A,t}\big(1 - L_{A,t}\big) - \delta_J a_{J,t} - \delta_S a_{S,t} - \kappa_{r\to L_A} r_t].$$

Within the agency update, legitimacy is computed from the incoming remuneration $\rho_t$, whereas the budget drain uses the updated $\rho_{t+1}$; the AI-operating-cost term uses the incoming $\varphi_t$ (one-step-lagged cost accounting).

*Capacity-responsive AI adoption:* The Agency estimates the level of AI-assisted review required to close the gap between proposal load and available human review capacity. Proposal load is defined as:

$Q_t = 1 + \chi p_t$, and

human review capacity is defined as:

$H_t$=$1 - r_t$.

The unconstrained requirement and its technically clipped counterpart are:

$$\varphi_t^{req,raw} = \frac{Q_t - H_t}{\beta_{AI}} = \frac{r_t + \chi p_t}{\beta_{AI}},$$

$$\varphi_t^{req} = \text{clip}_{[0,\varphi_{max}]}(\varphi_t^{req,raw}),$$

where $\varphi_t^{req,raw}$ is reported as a diagnostic but does not itself drive deployment.

Fiscal availability and the fiscally feasible target are:

$$a_M(B_t) = \text{clip}_{[0,1]}\left[\frac{B_t - B_{floor}}{1 - B_{floor}}\right],$$

$$\varphi_t^{target} = a_M(B_t)\, \varphi_t^{req}.$$

Deployment relaxes toward this feasible target at rate $r_\varphi$, while coordinated litigation above threshold $\tau_L$ rolls existing deployment back in proportion to the deployed level:

$$\varphi_{t+1} = \text{clip}_{[0,\varphi_{max}]}\left[\varphi_t + r_\varphi\left(\varphi_t^{target} - \varphi_t\right) - \delta_{\varphi S}([a_{S,t} - \tau_L]_+)\varphi_t\right].$$

Since $a_M(B_t) \to 0$ as the budget approaches its floor, deployment is suppressed by fiscal scarcity independently of the technical ceiling. By contrast, at baseline the budget is full ($a_M = 1$) and the target equals $\varphi_t^{req}$.

Reviewer remuneration, which might increase to provide incentives, tracks a resistance-responsive target so long as the budget remains above a floor, otherwise is reset to baseline:

$$\rho_t^* = \rho_{base} + r_t(\rho_{max} - \rho_{base}).$$

That is, if $B_t < B_{floor}$, then $\rho_{t+1} = \rho_{base}$; otherwise $\rho_{t+1} = \text{clip}_{0,\rho_{max}}[\rho_t + r_\rho(\rho_t^* - \rho_t)]$.

The budget stock is depleted by above-baseline pay and by AI operating costs, and replenished by exogenous regeneration:

$$B_{t+1} = \text{clip}_{[0,1]}\left[B_t - \delta_{drain}[\rho_{t+1} - \rho_{base}]_+ - q_\varphi \varphi_t + \delta_{regen}\right].$$

The pay drain uses the updated remuneration $\rho_{t+1}$, whereas the AI-operating-cost term uses the current deployment $\varphi_t$, following the model's one-step-lagged cost accounting.

All coefficients with their definitions, default values of the parameters, and calibration priorities are listed in the supplementary material. The derivation below uses these same equations and defines the threshold relations that the simulations subsequently test.

**Simulation parameters and stochasticity**

We adopted the functional form $\eta = 1 - 0.85c_{conv}$, because across $\eta$-slope 0.50→1.00 (a doubling), the cascade's terminal state is essentially unchanged (resist = 1.000, $L$ = 0, $\rho \approx 0.44$, budget ≈ 0.255, $K \approx 24.4$ throughout).

*Parameter Sweeps, bounding constraints, and finite-population stochasticity*: All simulations were implemented in R (script Evo_Dynamics_AI_Revision6_Master.R, seed 20240601), which regenerates every figure (for the appropriate figure renumbering, see text), the combined sweep_results_all.csv, and all diagnostic tables. Deterministic trajectories were advanced by the bounded Euler scheme described above: at each time unit, strategy frequencies within the Scientist and Public populations were incremented by the pairwise-comparison (Fermi) imitation flows, clipped to the unit interval, and renormalized onto their respective simplexes, while the five agency scalars ($L$, $L_A$, $\varphi$, $\rho$, $B$) were updated by their governing rules and clipped to their admissible ranges — the unit interval for $L$, $L_A$, and $B$, with $\varphi$ additionally bounded above by $\varphi_{max}$ and $\rho$ by $\rho_{max}$, and $\rho$ reset to $\rho_{base}$ whenever the budget fell below its floor. Local stability of a converged rest point $z*$ was assessed by the spectral radius of a one-sided finite-difference Jacobian (forward differences, $\varepsilon = 10^{-6}$). Because the baseline rest point places $L = 0$ and $L_A = 0$ on the clip boundary, the update map is non-smooth there; this Jacobian is therefore reported as a numerical diagnostic rather than a formal proof of stability, and perturbing simplex coordinates before renormalization mixes on- and off-simplex directions.

Each parameter sweep varied one pair of parameters over an 11 × 11 grid while holding all other parameters at their default values. For every grid cell the terminal value of the outcome variable was computed as a tail mean over the final segment of the trajectory rather than as a single endpoint, so that limit cycles are averaged over their oscillation; cells whose tail amplitude exceeded a threshold were flagged as unsettled or limit-cycling and are reported as such in the figure legends rather than silently averaged away. The verdict attached to each sweep (a wide, well-identified transition versus a graded effect) was generated in code from the measured span of the outcome variable across the grid.

Finite-population stochasticity was implemented as a Wright–Fisher sampling layer applied to the deterministic imitation dynamics: after each deterministic update, the Scientist and Public strategy vectors were resampled as multinomial draws of population size $n$ and renormalized. Its diffusion strength depends jointly on $n$, the integration step sizes, and the resampling interval. Therefore, the reported basin frequencies are frequencies under this specified stochastic algorithm. To assess robustness of this choice, we additionally implemented an explicit individual-level pairwise-comparison (Fermi) process, in which each agent independently revises its strategy with a per-step probability proportional to the integration step and copies a randomly drawn co-population member with the standard Fermi probability. This latter process's mean-field limit is identical to the deterministic kernel above, and its demographic noise per unit time-step is invariant to the integration scale. The sensitivity suite (Fig. 4D) applied both protocols across a grid of population sizes ($n$ = 100, 400, 1,600), resampling intervals (every 1, 5, or 25 steps), and integration scales ($h$ = 1, 0.5, 0.25), with 30 replicates per cell. For the metastability analysis we used $n$ = 400 with 100 replicate runs at baseline parameters, differing only in the realization of this sampling noise; terminal outcomes are reported as replicate means with 95% confidence intervals

**Global parameter-space sampling**

To test whether the regimes identified from targeted scenarios and pairwise sweeps organize the model's behavior globally, we drew a Latin-hypercube sample of $n$ = 5,000 points (seed 20240603) jointly varying 35 model parameters and two initial-condition components ($r_0$, the Scientist initial resisting fraction, and $p_0$, the Scientist initial proposal-intensifying fraction, with litigation as the residuals, over ranges given in the supplementary material; the Public initial condition was held at its default value, (0.65, 0.20, 0.15), throughout. Each dimension was drawn from an independent stratified design, with strata independently permuted across dimensions; $\rho_{max}$ was constrained to exceed $\rho_{base}$ by at least 0.05 wherever a draw violated this. Sampled $r_0$ and $p_0$ were mapped onto the four-strategy Scientist initial condition by weighting the pure refusal and proposal-intensification strategies more heavily than their combination, $sci_0 \propto (1 - r_0 - p_0, 0.7r_0, 0.7p_0, 0.15(r_0 + p_0))$, renormalized onto the simplex. Each draw was run to its adaptive terminus using the same checkpoint-based regime classification described above (2,000, 5,000, and subsequent 5,000-step checkpoints; maximum 20,000 time steps). All 5,000 draws completed successfully.

Terminal outcome variables (budget, effective capacity, fiscal availability, deployed AI, institutional and AI-review legitimacy, resistance, the residual capacity deficit, proposal load, and log-transformed congestion) were centered, scaled to unit variance, and reduced by principal component analysis. A Random-Forest classifier (500 trees, seed 20240604) was trained to predict the terminal two-axis regime (joint survival, capacity-only zombie state, legitimacy recovery without capacity, joint failure, or intermediate) from the 37 input dimensions; out-of-bag accuracy and permutation-based (mean-decrease-in-accuracy) variable importances are reported in Results.

**Use of AI Tools**

Generative AI systems: Claude (Anthropic; Claude Sonnet 4.5, Claude Sonnet 5, Claude Fable 5), ChatGPT (OpenAI; GPT5.5, GPT5.6), and Gemini (Google; Gemini 3.1Pro]) were used during preparation of this manuscript to identify errors in the model equations and their code implementation, to help design, debug, and extend the R simulation code (including the capacity-responsive AI-adoption controller and the two-axis outcome-classification scheme), and to draft, cross-check and revise text. All equations, parameter values, classification thresholds, and numerical results reported here were verified by the author by executing the deposited simulation code; no reported figure or statistic was taken directly from AI output without independent verification. The author is solely responsible for, and has fact-checked, all scientific content and conclusions in this work.

**Acknowledgement**

The initial formulation of the game-theoretic model was the author's own work. The author thanks NSF grant #0941078 (2009–2011) for supporting his earlier studies in evolutionary game theory; however, this specific study received no funding from any agency. Use of generative AI tools in preparing this manuscript is described in *Methods*. While writing this manuscript, the author was the W. M. Keck Foundation Professor of Systems, Computational and Molecular Biology at the Henry E. Riggs School of Applied Biological Sciences, Keck Graduate Institute, and a Visiting Faculty Associate at the Division of Biology and Biological Engineering, California Institute of Technology.

**APPENDIX I**

**Analytical Formulation of the Model**

*State Variables and Notations:* The system state is defined by the proportions of the populations utilizing each strategy, alongside the Agency's internal scalar variables. Let $\mathcal{S} = \{N, R, P, B\}$ denote the set of Scientist strategies, and $\mathcal{P} = \{A, J, S\}$ denote the set of Public strategies. The population state vectors at time $t$ are given by $f_t$ and $a_t$ over the standard simplex:

$$f_t = (f_{N,t}, f_{R,t}, f_{P,t}, f_{B,t}), \qquad a_t = (a_{A,t}, a_{J,t}, a_{S,t})$$

$$\sum_{i \in \mathcal{S}} f_{i,t} = 1, \sum_{m \in \mathcal{P}} a_{m,t} = 1, \ f_{i,t} \geq 0 \,, a_{m,t} \geq 0$$

The Agency's institutional state is governed by five scalar variables, all bounded to the interval [0,1]: institutional legitimacy $L_t$ , AI-review legitimacy $L_{A,t}$, AI adoption $\varphi_t$, reviewer remuneration $\rho_t$, and budget stock $B_t$. The complete state is defined by the vector:

$z_t = (f_t, a_t, L_t, L_{A,t}, \varphi_t, \rho_t, B_t)$

*Derived Operational Quantities*: To map strategic frequencies to systemic pressures, we define scientist-level indicator variables for resistance to review ($RES_i$), proposal intensification ($PROP_i$), and active reviewing ($REV_i$):

$RES_i = 1$, if $i \in \{R, B\}$, otherwise 0.

$PROP_i = 1$, if $i \in \{P, B\}$, otherwise 0.

$REV_i = 1$, if $i \in \{N, P\}$, otherwise 0.

These translate into aggregate systemic variables representing the total resistance fraction ($r_t$) and the proposal-intensification fraction ($p_t$):

$r_t$=$f_{R,t} + f_{B,t}$

$p_t$=$f_{P,t} + f_{B,t}$

We define the proposal load as $Q_t = 1 + \chi p_t$ , where the baseline proposal load is normalized to one, and $\chi$ measures the additional load generated by proposal intensification.

The overall effective review capacity ($C_t$) is a composite of the human review capacity ($H_t$=$1 - r_t$) and the supplementary capacity provided by AI ($\beta_{AI} \, \varphi_t$), where $\beta_{AI}$ represents the maximum capacity supplied by full AI adoption:

$$C_t = H_t + \beta_{AI} \, \varphi_t = 1 - r_t + \beta_{AI} \, \varphi_t$$

Finally, systemic congestion ($K_t$) is defined as the ratio of proposal load (scaled by a conversion factor $\chi$) to the effective capacity:

$K_t = \frac{(1+\chi\, p_t)}{\max\,(C_t,\varepsilon)}$, where $\varepsilon > 0$ prevents division by zero.

Thus, we derive, $\phi_t^{req} = \text{clip}_{[o,\phi_{max}]}\left[\frac{[Q_t - H_t]_+}{\beta_{\text{AI}}}\right]$, where $z_+ = \max\,(\text{z}, 0)$.

Since, $Q_t - H_t = (1+\chi p_t) - (\,1 - r_t) = r_t + \ \chi\, p_t$, we can rewrite:

$\phi_t^{req} = \text{clip}_{[o,\phi_{max}]}\left[\frac{r_t + \chi\, p_t}{\beta_{\text{AI}}}\right]$.

This is the minimum level of AI deployment needed to replace review capacity lost through refusal and to process the additional load caused by proposal load intensification.

We now define a fiscal availability factor:

$a_M(B_t) = \text{clip}_{[0,1]}\left[\frac{\left[B_t - B_{floor}\right]_+}{1 - B_{floor}}\right]$.

Thus, $a_M = 1$ when the budget stock is full.

$0 < a_M < 1$ when the budget is above the floor but is constrained,

$a_M = 0$ at or below the budget floor.

The fiscally feasible AI target is:

$\phi_t^{target} = a_M(B_t)\phi_t^{req}$.

The disruption term used in the institutional-legitimacy update is:

$$D_t = \text{clip}_{[0,1]}\big(r_t + a_{J,t} + a_{S,t}\big),$$

which makes visible the increases in disruption with scientist resistance, public rejection, and litigation. The public coordination multiplier for litigation is:

$$s_t = S\big(a_{J,t}, r_t, G\big) = \text{clip}_{[0,1]}\big(w_{S,0} + w_{S,J}a_{J,t} + w_{S,r}r_t + w_{S,G}G\big),$$

where $G \in [0,1]$ is the exogenous politicization level of the funding agency. The term $c_{conv} \in [0,1]$ denotes the fraction of scientists whose refusal is conviction-driven and relatively insensitive to payment. This value may be dependent on the career status of the scientists, but we neglect this factor in our simulation.

The main parameters are as follows: $c_0$ is the baseline scientist career payoff; $c_R$ is the career cost of refusing to review; $c_J$ and $c_S$ are public costs of rejection and litigation; $\omega_{prop}$ is the private value of extra proposal submission; $\omega_R$ is the collective-action benefit of resistance above the threshold $\tau_R$; $\omega_{norm}$ is the conformity or visibility benefit of joining a growing resistance; $\omega_{pay}$ is the weight placed on reviewer pay; $\omega_K$ is the congestion penalty; and $\omega_G$ and $\omega_L$ are the weights of politicization and legitimacy in scientist payoffs, respectively.

Public payoff weights are denoted by $w$ with subscripts identifying the actor and object being valued. Where the same actor–object pair is used both by the litigation-coordination multiplier and directly in the litigation payoff, the latter is distinguished with a prime ($w'$). The legitimacy-update parameters are denoted by $\kappa, \alpha, \delta$ and $\beta_{bribe}$, as defined in the equations below.

*Payoffs:* (a) Scientist payoffs combine career value, resistance costs and benefits, reviewer remuneration, proposal-congestion costs, politicization, and legitimacy. While the career value of accepting reviewer invitation is positive for an early career scientist, it is of low value for an established scientist; we neglect this subtlety. For a scientist strategy $i \in \mathcal{S}$, the scientist payoff,

$\pi_i^{(s)} = c_0 + \omega_{prop} PROP_i - c_R RES_i + \omega_R [r_t - \tau_R]_+ RES_i + \omega_{norm} r_t RES_i + \eta \rho_t \omega_{pay} REV_i - \omega_K PROP_i [K_t - 1]_+ + (-\omega_G G + \omega_L L_t) REV_i$ ---- **(Eq 1)**

(b) Public payoffs are:

$\pi_A^{(p)} = w_{A,L_A} L_{A,t} + w_{A,L} L_t - w_{A,\varphi} (1 - L_{A,t}) \varphi_t - w_{A,G} G$ ---- **(Eq 2)**

$\pi_J^{(p)} = w_{J,grow} (1 - L_{A,t}) \varphi_t + w_{J,G} G + w_{J,r} r_t - c_J$ ---- **(Eq 3)**

$\pi_S^{(p)} = s_t (w'_{S,G} G + w'_{S,\varphi} \varphi_t) + w'_{S,J} a_{J,t} - c_S$ ---- **(Eq 4)**

Public acceptance (high public trust) is favored by high institutional legitimacy and potentially also by high AI-review legitimacy, once the latter is established. Obviously, public rejection (low public trust) is favored by low AI-review legitimacy, high AI adoption under low legitimacy, politicization, and visible scientist resistance. Litigation is favored by public coordination, politicization, AI adoption, and prior rejection, but it is discouraged by its individual cost.

*Fermi Imitation Dynamics:* Within each strategic population, strategies spread by pairwise comparison using the Fermi imitation model in evolutionary dynamics (15). If an actor using strategy $i$ compares itself with an actor using strategy $j$, the probability of copying $j$ is:

$p(i \leftarrow j) = \sigma[\lambda(\pi_j - \pi_i)]$, where:

$\sigma(u) = \frac{1}{(1+e^{-u})}$, is the Fermi distribution for $u = \lambda(\pi_j - \pi_i)$.

In the large-population limit, the mean-field change in the frequency of scientist strategy $i$ is:

$\dot{f}_i = f_i \sum_{j \in \mathcal{S}} f_j \{ \sigma[\lambda_s (\pi_i^{(s)} - \pi_j^{(s)}) - \sigma[\lambda_s (\pi_j^{(s)} - \pi_i^{(s)})]\}$,

or equivalently it can be shown that,

$\dot{f}_i = f_i \sum_{j \in \mathcal{S}} f_j \, tanh[\frac{\lambda_s (\pi_i^{(s)} - \pi_j^{(s)})}{2}]$ ------ **(Eq 5)**

The public dynamics have the analogous form:

$\dot{a}_m = a_m \sum_{n \in \mathcal{P}} a_n \, tanh[\frac{\lambda_p (\pi_m^{(p)} - \pi_n^{(p)})}{2}]$ ----- **(Eq 6)**

For weak selection, *i.e.*, small payoff, tanh $(u) \approx u$, giving the replicator approximations:

$$\dot{f}_i \approx \left(\frac{\lambda_s}{2}\right) f_i (\pi_i^{(s)} - \bar{\pi}^{(s)})$$

$$\dot{a}_m \approx \left(\frac{\lambda_p}{2}\right) a_m (\pi_m^{(p)} - \bar{\pi}^{(p)})$$

With, $\bar{\pi}^{(s)} = \sum_{i \in s} f_i \, \pi_i^{(s)}$ and $\bar{\pi}^{(p)} = \sum_{n \in \mathcal{P}} a_m \, \pi_m^{(p)}$.

*Capacity and Congestion:* Operational throughput is feasible when proposal load does not exceed effective capacity $K_t \leq 1$. Using the definition of $K_t$, this condition is equivalent to $1 - r_t + \beta_{AI}\varphi_t \geq 1 + \chi p_t$.

Since this inequality contains the capacity variables $r_t$, $p_t$, and $\varphi_t$ but not the legitimacy variables $L_t$ or $L_{A,t}$, it formalizes the separation between capacity and certification. In other words, increasing AI adoption can independently restore processing capacity without repairing either form of legitimacy.

*Resistance Threshold*: The payoff gain from switching from a review-performing strategy to the corresponding refusal strategy (*i.e.*, resistance) is given by $\Delta_{res} = \pi_R^{(S)} - \pi_N^{(S)} = \pi_B^{(S)} - \pi_P^{(S)}$.

Substituting the scientist payoff equation (1) gives:

$\Delta_{res} = -c_R + \omega_R[r_t - \tau_R]_+ + \omega_{norm} r_t - \eta\rho_t\omega_{pay} + \omega_G G - \omega_L L_t$.

Resistance grows when $\Delta_{res} > 0$ and decays when $\Delta_{res} < 0$, when the refuse and proposal intensity traits are statistically independent in the population (linkage equilibrium). Below the collective-action threshold $r_t < \tau_R$, the collective-action term is absent, so resistance grows only if:

$\omega_{norm} r_t > c_R + \eta\rho_t\omega_{pay} - \omega_G G + \omega_L L_t$.

Above the threshold, $r_t > \tau_R$, the critical resistance fraction is:

$r_t > r_{crit} = [c_R + \eta\rho_t\omega_{pay} - \omega_G G + \omega_L L_t + \omega_R\tau_R]/(\omega_R + \omega_{norm})$.

Higher politicization lowers $r_{crit}$; higher refusal cost, reviewer pay, and institutional legitimacy raise it. Once $r_t$ exceeds both $\tau_R$ and $r_{crit}$, refusal becomes self-reinforcing under the imitation dynamics.

*Proposal Intensification Threshold:* The payoff gain from increased proposal submission, holding the review/refusal dimension fixed, is

$\Delta_{prop} = \pi_P^{(S)} - \pi_N^{(S)} = \pi_B^{(S)} - \pi_R^{(S)} = \omega_{prop} - \omega_K[K_t - 1]_+$.

If $K_t < 1$, proposal congestion is absent, and proposal intensification is favored whenever $\omega_{prop} > 0$. If $K_t > 1$, proposal intensification expands when

$\omega_{prop} > \omega_K[(1 + \chi p_t)/(1 - r_t + \beta_{AI}\varphi_t) - 1]$.

Equivalently, when congestion is active, proposal growth is favored below the implicit boundary

$p_t < \frac{\{[1+(\omega_{prop}/\omega_K)](1-r_t+\beta_{AI}\varphi_t)-1\}}{\chi}$.

Automation raises this boundary by increasing the effective capacity; by contrast, resistance lowers it by reducing human review capacity. This is the analytical argument for how proposal intensification can coexist with high AI adoption in the numerical trajectories.

*Public Acceptance, rejection or litigation:* The public subgame is governed by payoff differences. Acceptance dominates rejection when

$\Delta_{AJ} = \pi_A^{(p)} - \pi_J^{(p)} > 0$.

Substitution gives:

$$\Delta_{AJ} = w_{AL_A}L_{A,t} + w_{A,L}L_t - w_{A,\varphi}(1 - L_{A,t})\varphi_t - w_{A,G}G - w_{J,grow}(1 - L_{A,t})\varphi_t - w_{J,G}G - w_{J,r}r_t + c_J.$$

Acceptance dominates litigation when $\Delta_{AS} = \pi_A^{(p)} - \pi_S^{(p)} > 0$,

Or

$$\Delta_{AS} = w_{AL_A}L_{A,t} + w_{A,L}L_t - w_{A,\varphi}(1 - L_{A,t})\varphi_t - w_{A,G}G - s_t(w'_{S,G}G + w'_{S,\varphi}\varphi_t) - w'_{S,J}a_{J,t} + c_S.$$

Thus, acceptance is favored by high $L_t$ and $L_{A,t}$ and by high rejection or litigation costs. Rejection and litigation are favored by low $L_{A,t}$, high automation under low legitimacy, high politicization, visible scientist resistance, and coordinated public rejection.

*Legitimacy Rest-points:* For fixed $r, a_A$, $a_J$ and $a_S$, the AI-review legitimacy update has the following interior balance condition:

$$\alpha_A a_A(1 - L_A^*) - \delta_J a_J - \delta_S a_S - \kappa_{r\to L_A} r = 0.$$

Solving for $L_A^*$ gives:

$L_A^* = 1 - \frac{\delta_J a_J + \delta_S a_S + \kappa_{r\to L_A} r}{\alpha_A a_A}$, for $a_A > 0$.

AI-review legitimacy has a positive interior rest point only if $\alpha_A a_A > \delta_J a_J + \delta_S a_S + \kappa_{r\to L_A} r$.

If $a_A = 0$, or if this inequality fails, $L_A$ is driven toward the lower boundary. This is the analytical basis for the acceptance-gain versus rejection-damage phase diagram.

Institutional legitimacy has the analogous balance condition:

$$\kappa_{rep}(1 - L^*)a_A(1 - r) - \kappa_{pol}DG - \kappa_{rej}a_J - \beta_{bribe}[\rho - \rho_{base}]_+G = 0.$$

Solving for $L^*$ gives:

$$L^* = 1 - \frac{\kappa_{pol}DG + \kappa_{rej}a_J + \beta_{bribe}[\rho - \rho_{base}]_+G}{\kappa_{rep}a_A(1-r)}.$$

Institutional legitimacy survives only when repair by acceptance and participation exceeds politicization, rejection, and bribe-signal erosion:

$$\kappa_{rep}a_A(1 - r) > \kappa_{pol}DG + \kappa_{rej}a_J + \beta_{bribe}[\rho - \rho_{base}]_+G.$$

This condition shows that institutional legitimacy can collapse even when review capacity remains high. Capacity can be restored by $\varphi$, but institutional legitimacy requires acceptance $a_A$ and nonresistant participant $(1 - r)$.

*AI Adoption rollback and budget solvency:* Holding litigation fixed and ignoring boundary truncation, the AI adoption rest point satisfies:

$$0 = r_\varphi[a_M(B^*)\phi_{req}^* - \varphi^*] - \delta_{\varphi S}[a_S^* - \tau_L] + \varphi^*.$$

Therefore, $\varphi^* = \frac{r_\varphi a_M(B^*)\phi_{req}^*}{r_\varphi + \delta_{\varphi S}[a_S^* - \tau_L]_+}$.

For a budget-replete system without litigation above threshold, $B^* \approx 1$, and $a_S^* \leq \tau_L$ , the above equation reduces to $\varphi^* = \phi_{req}^* = \text{clip}_{[0,\varphi_{max}]}\left[\frac{r^* + \chi p^*}{\beta_{\text{AI}}}\right]$.

The reviewer pay tracks the resistance-responsive target, $\rho^* = \rho_{base} + r(\rho_{max} - \rho_{base})$ as long as $B \geq B_{floor}$. A necessary budget-solvency condition at the corresponding quasi-steady state is nonnegative budget drift:

$$\delta_{regen} \geq \delta_{drain}[\rho^* - \rho_{base}]_+ + q_\varphi \varphi^*.$$

Substituting the pay target gives:

$$\delta_{regen} \geq \delta_{drain} r(\rho_{max} - \rho_{base}) + q_\varphi \varphi^*.$$

When this inequality fails, the budget is pulled below its floor, pay is reset toward $\rho_{base}$, resistance may rise again, and the system can enter the relaxation oscillations observed near the fiscal boundary.

*Local stability and the role of simulation:* Let $z_{t+1} = F(z_t)$ represent one full bounded update of the coupled system, where $F$ is the staggered (Gauss–Seidel) composition defined in Methods: the population blocks advance from the common time-$t$ state and the agency block then advances from the updated frequencies and the time-$t$ agency scalars. The Jacobian $\mathbf{J}(\text{z}^*) = \frac{\partial F}{\partial z}$ is evaluated at the fixed point $\text{z} = \text{z}^*$ of this composed map. Since $F$ is staggered (not synchronous), $\mathbf{J}$ incorporates the within-step ordering, and the spectral radius $\text{spr}[\mathbf{J}(\text{z}^*)]$ governs local stability of the implemented dynamics.

**Table S1. Complete glossary of state variables, parameters, and constants.**

*All dynamical variables are bounded to [0, 1]. Defaults are the values used throughout unless a parameter sweep states otherwise. Calibration priority (Cal.): VH = very high, H = high, M = medium, L = low, — = structural/fixed.*

**A. Dynamical state variables (the system state is this 12-vector)**

| Symbol | Code name | Definition |
|---|---|---|
| $f_N$, $f_R$, $f_P$, $f_B$ | N, R, P, Both | Scientist strategy frequencies: normal, refuse, proposal-intensify, both |
| $a_A$, $a_J$, $a_S$ | A, Reject, Sue | Public strategy frequencies: accept, reject, litigate |
| $L_t$ | L | Institutional legitimacy |
| $L_{A,t}$ | LA | AI-review legitimacy |
| $\varphi_t$ | phi | AI adoption (deployed level) |
| $\rho_t$ | rho | Reviewer remuneration |
| $B_t$ | budget | Budget stock |

**B. Derived operational quantities (deterministic functions of the state)**

| Symbol | Code name | Definition |
|---|---|---|
| $r_t$ | resist | Aggregate resistance fraction, $f_R + f_B$ |
| $p_t$ | proposal_intensification | Proposal-intensifying fraction, $f_P + f_B$ |
| $H_t$ | human_capacity | Human review capacity, $1 - r_t$ |
| $Q_t$ | proposal_load | Proposal load, $1 + \chi \cdot p_t$ |
| $C_t$ | effective_capacity | Effective capacity, $H_t + \beta_{AI} \cdot \varphi_t$ |
| $K_t$ | congestion | Congestion ratio, $Q_t / \max(C_t, \varepsilon)$ |
| $D_t^a$ | disruption | Aggregate disruption, $\mathrm{clip}_{[0,1]}(r_t + a_{J,t} + a_{S,t})$ |
| $s_t^a$ | successP | Litigation coordination multiplier, $\mathrm{clip}_{[0,1]}(w_{S,0} + w_{S,J}\, a_{J,t} + w_{S,r}\, r_t + w_{S,G}\, G)$ |
| — | capacity_gap | Pre-AI capacity gap, $Q_t - H_t$ |
| — | residual_capacity_deficit | Post-AI deficit, $[Q_t - C_t]_+$ |
| $\varphi_{req,t}$ | phi_req_raw, phi_req | Required AI deployment (unconstrained; clipped to $\varphi_{max}$) |

| Symbol | Code name | Definition |
|---|---|---|
| — | technical_shortfall | $[\varphi_{req,raw} - \varphi_{max}]_+$ |
| $a_M(B_t)$ | fiscal_availability | Fiscal availability factor, ∈ [0, 1] |
| — | phi_target | Fiscally feasible AI target, $a_M \cdot \varphi^{req}$ |

a $D_t$ and $s_t$ are added here for completeness; both are used in Appendix I and in the code (successP) but were not separately tabulated in the original glossary.

**C. Exogenous setting and population-update parameters**

| Symbol | Code name | Default | Definition | Cal. |
|---|---|---|---|---|
| $G$ | G | 0.55 | Exogenous politicization level | H |
| $\lambda_s$ | lambda_sci | 3.0 | Scientist imitation selection strength | M |
| $\lambda_p$ | lambda_pub | 2.5 | Public imitation selection strength | M |
| $h_s$ | dt_sci | 0.015 | Scientist Euler step size | — |
| $h_p$ | dt_pub | 0.020 | Public Euler step size | — |

**D. Scientist resistance, proposal-load, and payoff parameters**

| Symbol | Code name | Default | Definition | Cal. |
|---|---|---|---|---|
| $\tau_R$ | tau | 0.25 | Collective-action threshold | H |
| $c_R$ | careerCost | 0.30 | Career cost of refusing to review | H |
| $c_{\text{conv}}$ | conviction | 0.25 | Conviction-driven, pay-insensitive fraction | VH |
| $\chi$ | chi | 1.50 | Proposal-load multiplier | H |
| $c_0$ | career_base | 0.90 | Baseline scientist career payoff | L |
| $\omega_R$ | omega_R | 1.10 | Collective-action (resistance) benefit | H |
| $\omega_{\text{norm}}$ | omega_norm | 0.25 | Conformity / visibility benefit | H |
| $\omega_{\text{pay}}$ | omega_pay | 0.65 | Weight on reviewer pay | M |
| $\omega_{\text{prop}}$ | omega_prop | 0.20 | Proposal career value | M |
| $\omega_K$ | omega_K | 0.30 | Congestion penalty | M |
| $\omega_G$ | omega_G | 0.45 | Politicization penalty (compliance) | H |
| $\omega_L$ | omega_L | 0.20 | Legitimacy reward (compliance) | H |

| Symbol | Code name | Default | Definition | Cal. |
|---|---|---|---|---|
| $\eta$ (derived) | — | — | Pay responsiveness = $1 - 0.85 \cdot c_{conv}$ (damped by conviction) | — |

**E. Public payoff parameters**

| Symbol | Code name | Default | Definition | Cal. |
|---|---|---|---|---|
| $w_{A,LA}$ | wA_LA | 0.45 | Accept: value of AI-review legitimacy | M |
| $w_{A,L}$ | wA_L | 0.20 | Accept: value of institutional legitimacy | M |
| $w_{A,\varphi}$ | wA_phi | 0.35 | Accept: penalty on unlegitimized AI | M |
| $w_{A,G}$ | wA_G | 0.15 | Accept: penalty on politicization | M |
| $w_{J,grow}$ | wJ_grow | 0.50 | Reject: value of unlegitimized-AI growth | H |
| $w_{J,G}$ | wJ_G | 0.25 | Reject: value of politicization | H |
| $w_{J,r}$ | wJ_rsci | 0.15 | Reject: value of visible scientist resistance | H |
| $c_J$ | rejectCost | 0.18 | Public cost of rejection | H |

| Symbol | Code name | Default | Definition | Cal. |
|---|---|---|---|---|
| $w_{S,0}$ | wS_succ0 | 0.20 | Litigation-coordination intercept | M |
| $w_{S,J}$ | wS_succJ | 0.40 | Coordination: prior rejection | M |
| $w_{S,r}$ | wS_succR | 0.20 | Coordination: scientist resistance | M |
| $w_{S,G}$ | wS_succG | 0.10 | Coordination: politicization | M |
| $w'_{S,G}$ | wS_G | 0.40 | Litigate: value of politicization | M |
| $w'_{S,\varphi}$ | wS_phi | 0.30 | Litigate: value of AI adoption | M |
| $w'_{S,J}$ | wS_J | 0.20 | Litigate: value of prior rejection | M |
| $c_S$ | suitCost | 0.25 | Public cost of litigation | M |

Note: unprimed $w_{S,*}$ parameterize the litigation-coordination multiplier $s_t$; primed $w'_{S,*}$ appear directly in the litigation payoff (Appendix I, Eq. 4).

**F. AI-adoption controller and litigation rollback**

| Symbol | Code name | Default | Definition | Cal. |
|---|---|---|---|---|
| $\beta_{AI}$ | beta_AI | 0.70 | AI capacity multiplier (capacity per unit $\varphi$) | M |
| $\varphi_{max}$ | phi_max | 0.85 | AI deployment ceiling | M |

| Symbol | Code name | Default | Definition | Cal. |
|---|---|---|---|---|
| $r_{\varphi}$ | r_phi | 0.012 | AI adoption adjustment rate | M |
| $\delta_{\varphi S}$ | phi_rollback | 0.040 | Litigation-driven adoption rollback | M |
| $\tau_L$ | tau_L | 0.18 | Litigation rollback threshold | M |
| $q_{\varphi}$ | q_phi | 0.015 | Budget cost per unit AI adoption | L |

**G. Legitimacy dynamics**

| Symbol | Code name | Default | Definition | Cal. |
|---|---|---|---|---|
| $\kappa_{\text{rep}}$ | kappa_rep | 0.035 | Institutional-legitimacy repair rate | VH |
| $\kappa_{\text{pol}}$ | kappa_pol | 0.060 | L loss from politicized disruption | H |
| $\kappa_{\text{rej}}$ | kappa_rej | 0.030 | L loss from rejection | H |
| $\beta_{\text{bribe}}$ | beta_bribe | 0.45 | L loss from emergency-pay bribe signal | VH |
| $\alpha_A$ | acc_gain | 0.055 | LA gain from acceptance | VH |
| $\delta_J$ | rej_damage | 0.050 | LA loss from rejection | H |

| Symbol | Code name | Default | Definition | Cal. |
|---|---|---|---|---|
| $\delta_S$ | legal_damage | 0.060 | LA loss from litigation | H |
| $\kappa_{\mathrm{r}\to\mathrm{LA}}$ | kappa_resist_LA | 0.030 | LA loss from scientist resistance | M |

**H. Reviewer remuneration and budget dynamics**

| Symbol | Code name | Default | Definition | Cal. |
|---|---|---|---|---|
| $\rho_{\mathrm{base}}$ | rho_base | 0.10 | Remuneration floor / baseline | M |
| $\rho_{\mathrm{max}}$ | rho_max | 0.80 | Remuneration ceiling | M |
| $r_\rho$ | r_rho | 0.06 | Remuneration adjustment rate | M |
| $\delta_{\mathrm{drain}}$ | delta_drain | 0.050 | Budget drain from above-baseline pay | M |
| $\delta_{\mathrm{regen}}$ | delta_regen | 0.018 | Budget regeneration | M |
| $B_{\mathrm{floor}}$ | budget_floor | 0.20 | Pay-suspension budget threshold | M |
| — | budget_pay_rule | hard | Pay-reset rule at floor (hard / smooth) | — |
| — | budget_reset_width | 0.02 | Smoothing width (smooth rule only) | — |

**I. Initial conditions and numerical constants**

| Symbol | Code name | Default | Definition |
|---|---|---|---|
| — | sci0 | 0.70, 0.15, 0.10, 0.05 | Initial (N, R, P, B) |
| — | pub0 | 0.65, 0.20, 0.15 | Initial (A, J, S) |
| $L_0$ | L0 | 0.80 | Initial institutional legitimacy |
| $L_{A,0}$ | LA0 | 0.60 | Initial AI-review legitimacy |
| $\varphi_0$ | phi0 | 0.05 | Initial AI adoption |
| $\rho_0$ | rho0 | 0.10 | Initial remuneration |
| $B_0$ | B0 | 0.80 | Initial budget stock |
| $\varepsilon$ | epsilon | $1 \times 10^{-6}$ | Division-by-zero safeguard |

**Table S2. Correspondence between analytical threshold conditions (Appendix I) and their numerical tests.**

*Each row states an analytical condition, its qualitative prediction for the coupled system, and the simulation results used to evaluate it. Conditions involving population frequencies are quasi-static analytical relations; the simulations determine whether they remain predictive under simultaneous feedback among Scientists, Public actors, and the Funding Agency.*

| Analytical result | Threshold or condition | Testable prediction | Revised numerical verification |
|---|---|---|---|
| Capacity–legitimacy separation and technical sufficiency | Neither $L_t$ nor $L_{A,t}$ enters this condition. Under unconstrained deployment, technical restoration additionally requires $r_t + \chi\, p_t \leq \beta_{AI}\, \varphi_{max}$. | Operational capacity and legitimacy can vary independently. AI can restore throughput only if its maximum attainable capacity is sufficient and deployment is not prevented by fiscal scarcity or litigation. | At the baseline fixed point, $\varphi^* = 0.85$ and $B^* = 1$, but $\varphi^*_{req,raw} = 2.143$, $K^* = 1.567$, and $L^* = L^*_A = 0$: both capacity and legitimacy fail. Under transparent rollout, $K^* = 1.567$ remains unchanged while $L^* = L^*_A = 1$, demonstrating legitimacy without capacity. Finite-population simulations additionally produce a capacity-only state in 16% of replicates. Figs. 2, 3, 5D, and 6; Supplementary Figs. S1, S4, S5. |
| Scientist-resistance cascade threshold | Resistance grows when $\Delta_{res} > 0$. Above the collective-action threshold $\tau_R$, $r_t > r_{crit} = (c_R + \eta\, \rho_t\, \omega_{pay} - \omega_G\, G + \omega_L\, L_t + \omega_R\, \tau_R) / (\omega_R + \omega_{norm})$. | Higher politicization lowers the resistance threshold. Higher refusal cost, effective reviewer remuneration, and institutional legitimacy raise it. Once $r_t$ exceeds both $\tau_R$ and $r_{crit}$, resistance becomes self-reinforcing and outcomes can depend on the initial resistance seed. | The career-cost–politicization sweep shows a sharp diagonal resistance frontier, with a narrow band of collapse persisting within the fixed region at high politicization. Selection strength, collective-action threshold, and conviction alter its position and sharpness. High initial resistance produces a resistance-cascade regime, whereas lower seeds lead to low-resistance outcomes. The cascade remains unresolved after 50,000 time steps rather than converging to a demonstrated limit cycle. Fig. 3; Fig. 5A; Supplementary Fig. S2. |

| Analytical result | Threshold or condition | Testable prediction | Revised numerical verification |
|---|---|---|---|
| Proposal-intensification boundary | $\Delta_{\text{prop}} = \omega_{\text{prop}} - \omega_K [K_t - 1]_+$. Proposal intensification is favored when $\Delta_{\text{prop}} > 0$. | AI deployment raises effective capacity and therefore delays the congestion penalty, allowing proposal intensification to coexist with high AI use. If its private payoff remains positive, proposal intensification can itself generate a capacity deficit that exceeds the maximum AI response. | At both the baseline and transparent-rollout fixed points, refusal approaches zero while proposal intensification approaches one. Proposal load reaches $Q^* = 2.50$, whereas effective capacity reaches only $C^* = 1.595$. Endogenous proposal intensification—not refusal—drives the deterministic baseline operational failure. Figs. 2 and 3; Supplementary Fig. S4. |
| Public subgame: acceptance, rejection, and litigation | Acceptance dominates rejection when $\Delta_{\text{AJ}} = \pi_A - \pi_J > 0$, and acceptance dominates litigation when $\Delta_{\text{AS}} = \pi_A - \pi_S > 0$. | Acceptance is favored by high $L_t$, high $L_{\text{A,t}}$, and high costs of rejection or litigation. Rejection is favored by low AI-review legitimacy, high unlegitimized AI deployment, politicization, and visible scientist resistance. Litigation is favored by coordination, politicization, AI deployment, prior rejection, and low litigation cost. | Public rejection approaches fixation in the baseline, whereas acceptance approaches fixation under transparent rollout. Litigation rises as both its individual cost $c_S$ and coordination threshold $\tau_L$ decline. Figs. 2 and 3; Fig. 5C. |
| AI-review legitimacy survival | An interior positive rest point requires $\alpha_A a_A > \delta_J a_J + \delta_S a_S + \kappa_{r\to LA}\, r$. | AI-review legitimacy survives only when legitimacy repair generated by acceptance exceeds erosion from rejection, litigation, and scientist resistance. Adoption speed alone does not determine legitimacy survival. | The acceptance-gain–rejection-damage sweep shows a sharp boundary between $L^*_A = 0$ and $L^*_A = 1$. Increasing acceptance gain alone changes the baseline legitimacy-collapse state into the transparent-rollout legitimacy-survival state even though deployment and congestion are essentially unchanged. Fig. 3; Fig. 5B. |
| Institutional-legitimacy survival | $\kappa_{\text{rep}}\, a_A\, (1\text{-}r) > \kappa_{\text{pol}}\, D\, G + \kappa_{\text{rej}}\, a_J + \beta_{\text{bribe}}\, [\rho - \rho_{\text{base}}]_+\, G$. | Institutional legitimacy survives when repair through public acceptance and | Baseline and transparent rollout have essentially identical operational states but opposite |

| Analytical result | Threshold or condition | Testable prediction | Revised numerical verification |
|---|---|---|---|
| | | nonresistant scientific participation exceeds politicized disruption, rejection, and bribe-signal erosion. High emergency remuneration damages legitimacy only when above-baseline pay and politicization coexist. | institutional-legitimacy fixed points, $L^* = 0$ and $L^* = 1$, respectively. Repair rate and politicization generate a sharp legitimacy boundary, though within the tested plane politicization alone selects the terminal basin; repair rate governs only transient speed. Bribe signaling becomes relevant in regimes in which resistance sustains remuneration above $\rho_{\text{base}}$. Figs. 2 and 3; Fig. 5C; Supplementary Fig. S3. |
| Capacity-responsive AI deployment and litigation rollback | The required deployment is $\varphi^{\text{req}}{}_t = \text{clip}_{[0,\varphi_\{\max\}]}\}((r_t + \chi\, p_t)/\beta_{\text{AI}})$. The interior rest point is $\varphi^* = (r_\varphi\, a_M(B^*)\, \varphi^*{}_{\text{req}}) / (r_\varphi + \delta_{\varphi S}[a^*{}_S - \tau_L]_+)$. | Deployment increases only in response to the capacity deficit, is reduced by fiscal scarcity, and is rolled back by litigation above $\tau_L$. Even full deployment cannot restore throughput when the unconstrained requirement exceeds $\varphi_{\max}$. | At baseline, fiscal availability is complete and litigation is negligible, so deployment reaches $\varphi_{\max} = 0.85$; nevertheless, the unconstrained requirement is 2.143 and congestion remains $K^* = 1.567$. No tested combination of $\beta_{\text{AI}}$ and $\varphi_{\max}$ reaches $K \leq 1$. In the resistance cascade, fiscal scarcity reduces deployment to a terminal-window mean of approximately 0.059. Figs. 2, 3, and 5D; Supplementary Fig. S4. |
| Budget-solvency condition | At quasi-steady state, a necessary solvency condition is $\delta_{\text{regen}} \geq \delta_{\text{drain}}\, r\, (\rho_{\max} - \rho_{\text{base}}) + q_\varphi\, \varphi^*$. | The relative importance of remuneration drain and AI operating cost depends on the dynamical regime. When resistance is low, remuneration returns to its baseline and AI operating cost dominates fiscal pressure. Under sustained resistance, above-baseline remuneration and low fiscal availability can jointly suppress deployment. | In the low-resistance baseline, the drain-by-regeneration sweep is dominated by horizontal regeneration bands rather than a diagonal remuneration-drain frontier. The AI-cost–regeneration sweep displays the clearer fiscal boundary. Under the hard budget-floor rule, the cascade remains persistently fluctuating and unresolved; smoothing the floor produces fixed points with severe operational collapse. The model supports a robust fiscal collapse but not a |

| Analytical result | Threshold or condition | Testable prediction | Revised numerical verification |
|---|---|---|---|
| | | | robust stable limit cycle. Fig. 5D; Supplementary Fig. S4, budget-floor-rule sensitivity analysis. |
| Finite-population basin selection† | Multinomial resampling produces sampling fluctuations of order $N^{-1/2}$ in Scientist and Public strategy frequencies. | Finite-population drift can move early trajectories across basin boundaries and select among alternative long-lived regimes. Continued switching after basin establishment is not required. | With $\mathrm{N} = 400$, 100 replicates, a 5,000-step burn-in, and a 5,000-step observation period, replicates partitioned into five outcomes under the two-axis (capacity × legitimacy) classification: 49% joint survival, 16% a capacity-only zombie state, 9% legitimacy recovery without capacity, 19% joint failure (68% of these resistance-driven), and 7% an intermediate class (entirely resistance-driven). No switching among basins was observed during the observation period. A dedicated sensitivity suite further shows this split is a property of the assumed stochastic algorithm, not a fixed probability: joint survival and the zombie state both decline toward single digits as population size grows and are largely absent under an individual-level pairwise-comparison process, while joint failure and legitimacy-without-capacity persist across every specification tested. Fig. 4A–B; Supplementary Fig. S5. |

*† Stochastic basin selection is a simulation discovery enabled by the threshold analysis: the analytics locate the basin boundary; finite-population drift across it is a purely numerical finding.*

## Supplementary Methods

All figures, tables, and numbers below are regenerated from Evo_Dynamics_AI_Revision6_Master.R.

**S1. Model specification pseudocode**

Three actor classes. Scientists choose N (normal participation), R (refuse to review), P (increased proposal submission), or B (both). Public actors choose A (accept AI review), J (reject publicly), or S (litigate). The agency is a non-imitating adaptive controller updating AI adoption (phi) and remuneration (rho) under a budget stock (B). State vector x = {f_N,f_R,f_P,f_B, a_A,a_J,a_S, L, L_A, phi, rho, B}; all continuous variables are bounded to [0,1].

Operational quantities (deterministic functions of the state):

```
r_sci   = f_R + f_B                 # resisting fraction
p_sub   = f_P + f_B                 # proposal-increasing fraction
H       = 1 - r_sci                 # human review capacity
Q       = 1 + chi * p_sub           # proposal load; chi = proposal-load multiplier (default 1.5)
C       = H + beta_AI * phi         # effective capacity
K       = Q / max(C, eps)           # congestion ratio

phi_req_raw = (Q - H) / beta_AI     # unconstrained requirement (reported)
phi_req     = clip(phi_req_raw, 0, phi_max)
a_M         = clip( (B - budget_floor)/(1 - budget_floor), 0, 1 )   # fiscal availability
phi_targ    = a_M * phi_req
```

Scientist payoffs: Only review-performing strategies (N, P) earn reviewer pay and the engagement-legitimacy term; refusers (R, B) do not, and proposal-increasers (P, B) carry the congestion penalty.

```
eta      = 1 - 0.85*conviction            # pay response, damped by conviction
benefit  = omega_R * max(r_sci - tau, 0)  # collective-action benefit
career_i = career_base + omega_prop*PROP_i - careerCost*RES_i
norm_i   = omega_norm * RES_i * r_sci
pay_i    = eta * rho * omega_pay * REV_i  # REV_i = 1 for N,P
cong_i   = - omega_K * PROP_i * max(K-1, 0)
legit_i  = (-omega_G*G + omega_L*L) * REV_i
pi_i     = career_i + benefit*RES_i + norm_i + pay_i + cong_i + legit_i
```

Public payoffs: Costs are flat; coordination enters only through bounded terms (no cost-to-zero runaway).

```
successP = clip(wS_succ0 + wS_succJ*a_J + wS_succR*r_sci + wS_succG*G, 0, 1)
pi_A = wA_LA*L_A + wA_L*L - wA_phi*(1-L_A)*phi - wA_G*G
pi_J = wJ_grow*(1-L_A)*phi + wJ_G*G + wJ_rsci*r_sci - rejectCost
pi_S = successP*(wS_G*G + wS_phi*phi) + wS_J*a_J - suitCost
```

Imitation (pairwise Fermi): (Optional finite-population multinomial resampling adds drift)

```
P(i copies j) = 1 / (1 + exp[-lambda*(pi_j - pi_i)])
df_i = f_i * sum_j f_j * [P(j copies i) - P(i copies j)]
# then Euler step + renormalize (same form for the Public population, using a_m and lambda_pub)
```

Agency updates: computed from the SAME time-t agency state (L, L_A, phi, rho, B) but the JUST-UPDATED population frequencies (f', a') -- a staggered (Gauss-Seidel), not synchronous, ordering. L is two-sided (repair vs erosion); L_A gain saturates; adoption rollback uses its own parameter; pay collapses to base when the budget falls below a floor.

```
D    = clip(r_sci + a_J + a_S, 0, 1)
L'   = clip(L + kappa_rep*(1-L)*a_A*(1-r_sci) - kappa_pol*D*G
            - kappa_rej*a_J - beta_bribe*max(rho-rho_base,0)*G, 0, 1)
L_A' = clip(L_A + acc_gain*a_A*(1-L_A) - rej_damage*a_J
            - legal_damage*a_S - kappa_resist_LA*r_sci, 0, 1)
phi' = clip(phi + r_phi*(phi_targ - phi)
            - phi_rollback*max(a_S - tau_L, 0)*phi, 0, phi_max)
rho* = rho_base + r_sci*(rho_max - rho_base)
rho' = rho_base                              if B < budget_floor
       clip(rho + r_rho*(rho* - rho), 0, rho_max)  otherwise
B'   = clip(B - delta_drain*max(rho' - rho_base, 0) - q_phi*phi + delta_regen, 0, 1)
```

## SUPPLEMENTARY RESULTS

**Supplementary Results Fig. S1:** Initial-condition dependence. At baseline parameters, three Scientist initial-resistance seeds reach three distinct outcomes: a low seed ($r_0$ = 0.07) reaches legitimacy without capacity (L* = L_A* = 1.0); an intermediate seed ($r_0$ = 0.20, the default) reaches joint failure with negligible resistance; and a high seed ($r_0$ = 0.45) reaches joint failure with resistance at fixation (φ* ≈ 0.06, budget* ≈ 0.25). The same deterministic-versus-resistance-driven duality shown stochastically by the Fig. 4 example, therefore also appears from initial conditions alone.

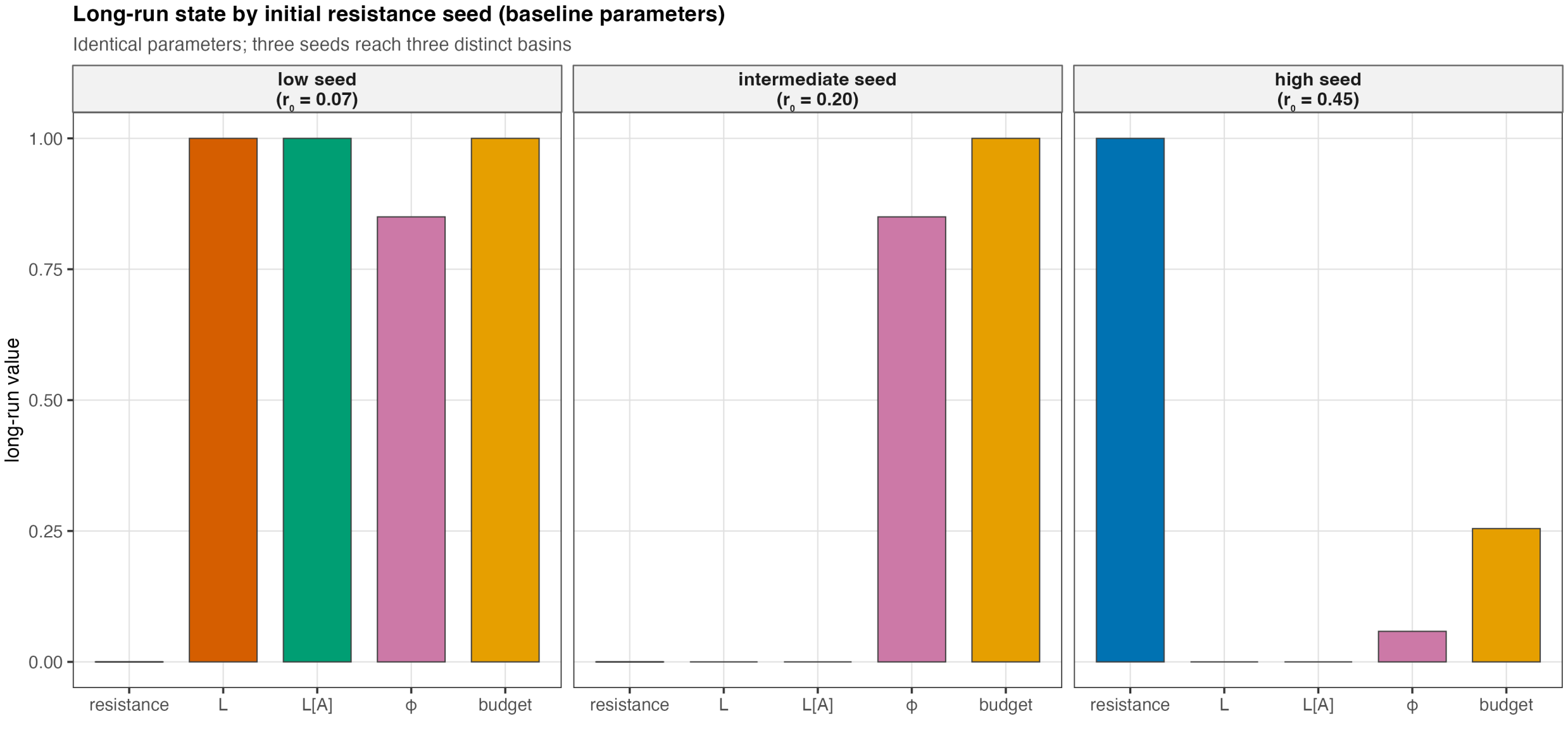

**Supplementary Results Fig. S2:** Additional resistance-frontier sweeps. Three further sweeps extend Fig. 5A's career-cost–politicization frontier. Selection strength ($\lambda$_sci, Fig. S2C) leaves the frontier's location essentially unchanged, consistent with its role in the replicator limit as a speed, not a sign, of the payoff-driven flow. The collective-action threshold ($\tau$_R, Fig. S2B) and the conviction-driven fraction (Fig. S2A) both shift it: higher $\tau$_R delays fixation to higher politicization, and higher conviction, a larger pay-insensitive fraction, pushes the frontier toward higher career costs, since remuneration becomes a weaker deterrent to resistance.

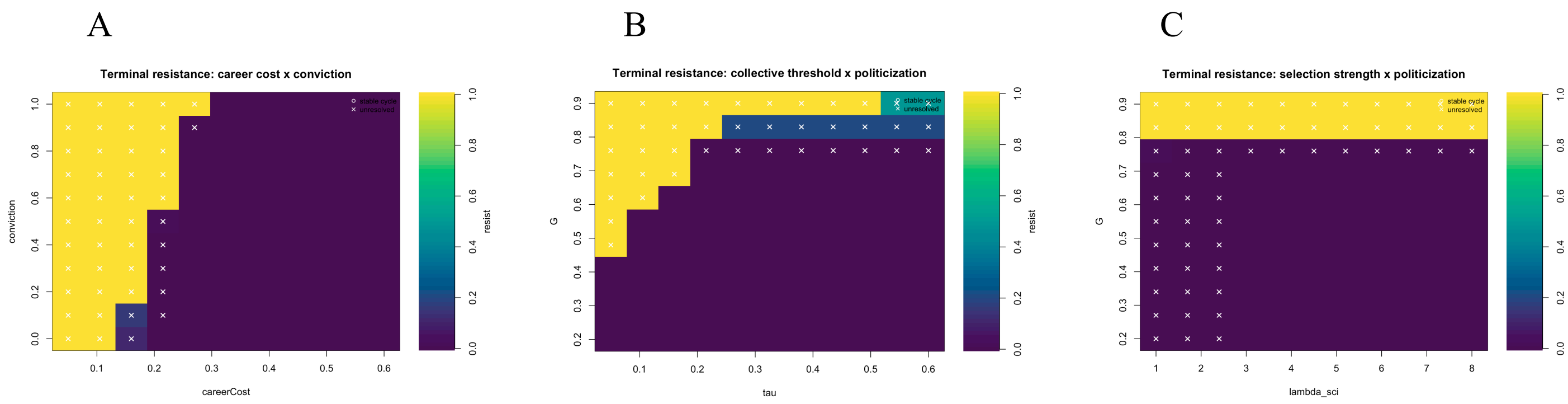

**Supplementary Results Fig. S3:** AI legitimacy-boundary sweeps. Four further sweeps extend the acceptance-gain and repair-rate boundaries of Fig. 5B–C. The bribe-signal coefficient (β_bribe, Fig. S3A) does not alter terminal institutional legitimacy in the low-resistance regime, since remuneration returns to baseline and the bribe term is inactive — the same G ≈ 0.5 threshold from Fig. 5C reappears unchanged. Litigation (Fig. S3B) responds in a graded, not threshold-like, manner to its individual cost and coordination threshold. AI-review legitimacy shows the same sharp step boundary against litigation damage (Fig. S3C) as against rejection damage in the main text, and higher public rejection costs lower the acceptance gain required for survival (Fig. S3D).

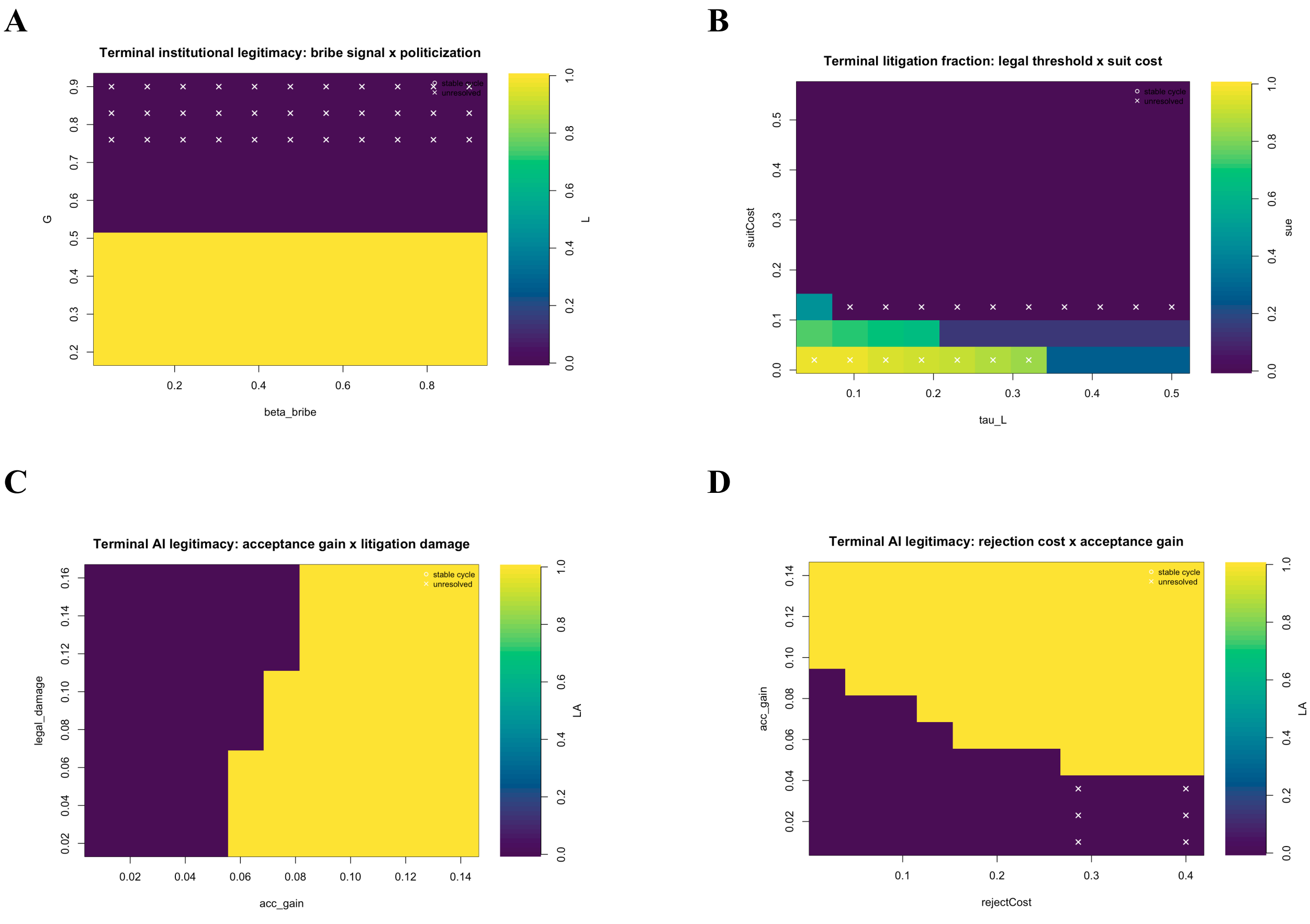

**Supplementary Results Fig. S4: Fiscal and capacity robustness.** Three further sweeps extend Fig. 5D. Budget solvency (Fig. S4A) is governed by regeneration relative to a threshold near $\delta_{regen} \approx 0.013$, essentially independent of the remuneration-drain coefficient over a tenfold range. Fiscal availability for AI (Fig. S4B) declines continuously with operating cost and rises with regeneration, exposing the same tradeoff more directly. The unconstrained AI requirement (Fig. S4C) ranges from approximately 0.08 to 4.75 across the tested plane and remains above the deployment ceiling everywhere tested. The closed-form sufficiency condition $\beta_{AI} \cdot \varphi_{max} \geq \chi$ is never satisfied within the explored range.

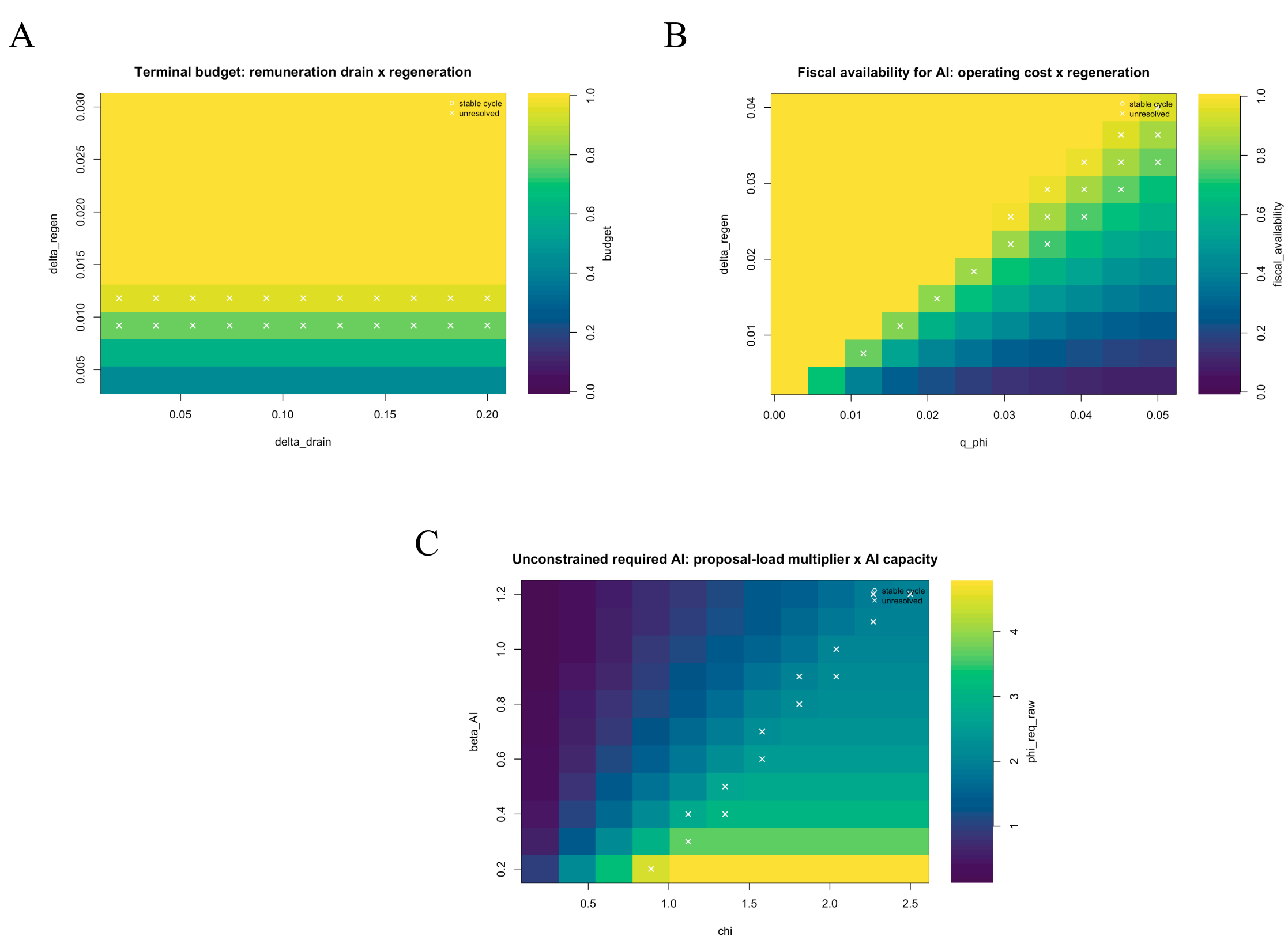

## Supplementary Results Fig. S5:

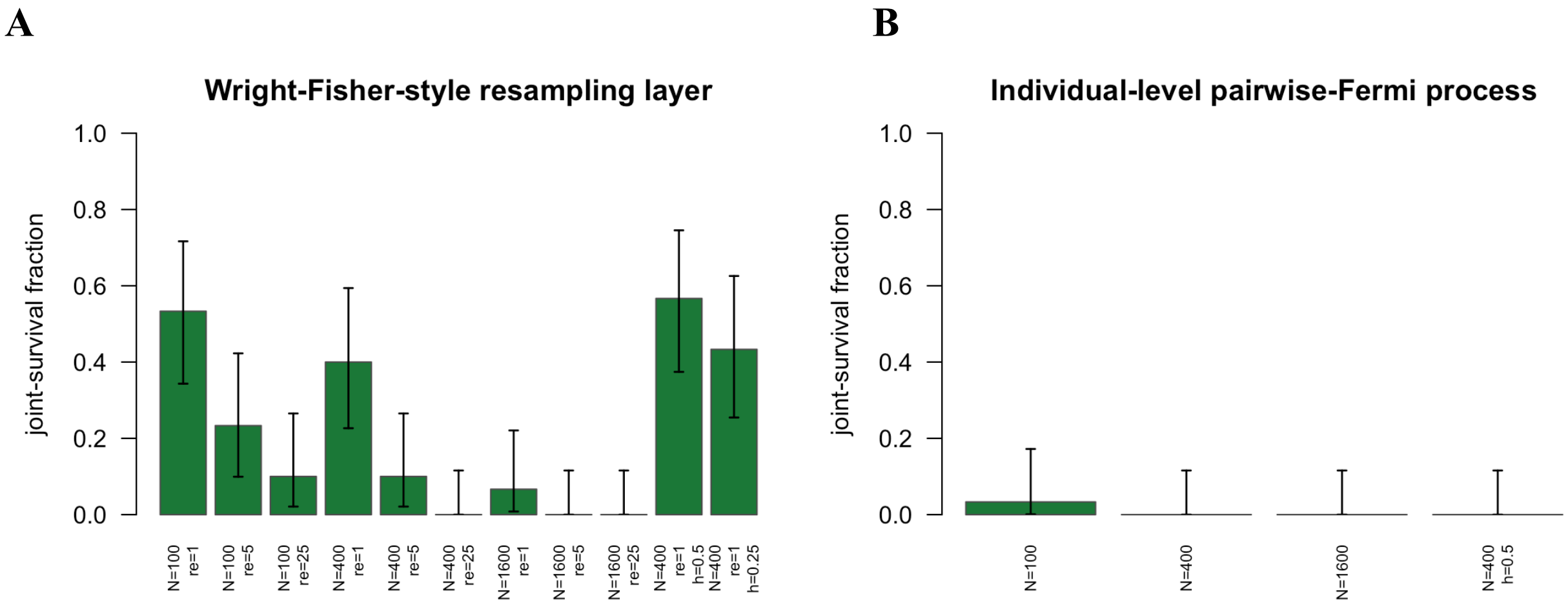

**Supplementary Results Fig. S6: Numerical robustness.** Two checks confirm the results are not artifacts of the numerical implementation (Fig. S6). Refining the integration step ($h$ = 1, 0.5, 0.25) leaves baseline and transparent-rollout congestion invariant to better than $10^{-8}$ relative, floating-point roundoff, not a resolvable numerical effect, while cascade congestion drifts by 0.96% (24.20 → 23.97), a small, monotonic change that does not affect its classification. Replacing the hard budget-floor pay reset with a smoothed version eliminates the cascade's persistent oscillation (all three smoothed widths converge to fixed points) but at the cost of far more severe capacity collapse: terminal congestion rises from $K^* \approx 24$ under the hard rule to $K^*$ = 513–2,395 under smoothing, 21 to 99 times higher. The qualitative finding — a resistance-driven joint failure — is robust to this implementation choice; its exact severity and settling behavior are not.

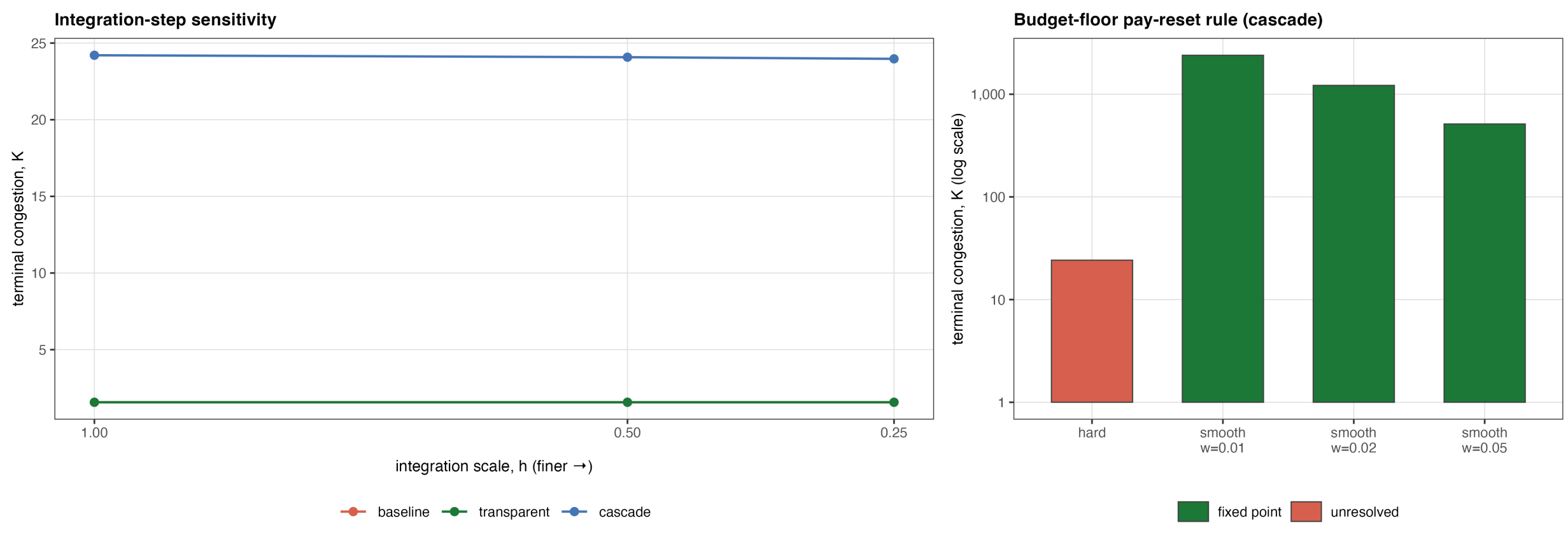

## S7. Reproducibility

1. Evo_Dynamics_AI_Revision6_Master.R is self-contained, sets seed 20240601, and writes all figures, the combined sweep_results_all.csv, and all diagnostic CSV files. Should be run on at least 12 cpu cores for a reasonable runtime.